\documentclass[trackchanges]{aastex701}
\usepackage{multirow}
\usepackage{graphicx}
\usepackage{booktabs}
\usepackage{makecell}
\usepackage{bm}
\usepackage{subfigure}
\usepackage{amsmath,amssymb}
\usepackage[T1]{fontenc}
\graphicspath{{./}{figures/}}
\usepackage{longtable}
\usepackage{cases}
\usepackage{epstopdf}
\usepackage[]{amsmath}
\usepackage{threeparttablex}
\usepackage{mathtools}
\usepackage{comment}
\usepackage{siunitx}

\begin{document}

\title{Constraints on neutrino emission and hadronic flux from 1LHAASO catalog $\gamma$-ray sources}

\author[orcid=0009-0007-5729-8062]{Xue-Rui Ouyang}
\affiliation{Laboratory for Relativistic Astrophysics, Department of Physics, Guangxi University, Nanning 530004, China}
\email{xr-ouyang@st.gxu.edu.cn}

\author[orcid=0000-0002-6316-1616]{Yun-Feng Liang}
\affiliation{Laboratory for Relativistic Astrophysics, Department of Physics, Guangxi University, Nanning 530004, China}
\email[show]{liangyf@gxu.edu.cn}

\author{Shi-Long Chen}
\affiliation{Laboratory for Relativistic Astrophysics, Department of Physics, Guangxi University, Nanning 530004, China}
\email{slchen@ihep.ac.cn}

\author[orcid=0009-0002-9585-0210]{Rong-Lan Li}
\affiliation{Key Laboratory of Dark Matter and Space Astronomy, Purple Mountain Observatory, Chinese Academy of Sciences, Nanjing 210023, China}
\email{rlli@pmo.ac.cn}

\author[orcid=0000-0002-9067-3828]{Ming-Xuan Lu}
\affiliation{Laboratory for Relativistic Astrophysics, Department of Physics, Guangxi University, Nanning 530004, China}
\email{lumx@st.gxu.edu.cn}

\begin{abstract}
IceCube has detected neutrino emission from the Galactic Plane (GP) at a significance of $4.5\sigma$, though its origin remains uncertain. Utilizing ten years of IceCube muon-track data, we investigate potential correlations between the GP neutrinos and $\gamma$-ray sources in the first LHAASO catalog (1LHAASO). To avoid issues caused by spectral extrapolation, this analysis focuses on sources detected by the Water Cherenkov Detector Array (WCDA). We employ an unbinned likelihood analysis to search for neutrino emission and constrain the hadronic $\gamma$-ray component of these sources. Neither single-source searches nor stacking analyses reveal significant neutrino signals. 
The stacking analysis indicates that the 1LHAASO WCDA population contributes at most $\sim$20\% to the diffuse GP neutrino flux measured by IceCube. The total hadronic contribution to the cumulative $\gamma$-ray emission from all WCDA sources is constrained to be at most $\sim$$60\%$, suggesting a predominantly leptonic origin for the $\gamma$-ray emission from the LHAASO source population. Even accounting for unresolved sources below the detection threshold, we estimate the total neutrino flux from all discrete sources (resolved plus unresolved) reaches at most about 40\% of the observed GP neutrino flux. These results support that the bulk of the GP neutrino emission is mainly from truly diffuse processes, i.e., cosmic-ray interactions with the interstellar medium, rather than from unresolved point sources.
\end{abstract}
\keywords{Neutrino astronomy (1100) --- Gamma-rays sources(633) --- High energy astrophysics (739)}


\section{Introduction}
Cosmic rays have been observed for over a century~\cite{Hess:1912srp}, with energies reaching up to $\sim 10^{20}$~eV~\cite{HIRES:1994ijd}. The origins of high-energy cosmic rays remain uncertain. As cosmic rays propagate through the interstellar medium, their trajectories are deflected by magnetic fields, making it difficult to identify their sources using directional information. However, cosmic rays may interact with gas or radiation fields surrounding their sources, producing charged and neutral pions that subsequently decay into neutrinos and gamma rays. Unlike cosmic rays, neutrinos and gamma rays propagate undeflected by magnetic fields and can thus serve as direct indicators of cosmic-ray sources. Below the knee ($\sim 4$~PeV) of the cosmic-ray spectrum, cosmic rays are generally believed to originate from within the Galaxy. Widely accepted candidates for Galactic cosmic-ray sources include supernova remnants (SNRs), pulsar wind nebulae (PWNs), and young massive star clusters (YMCs), yet direct observational evidence remains elusive.

Neutrinos provide an ideal messenger for astrophysical observations. In contrast to high-energy photons, which may interact with radiation fields or the interstellar medium and be absorbed during propagation to Earth, neutrinos interact only weakly with matter, allowing them to traverse dense media with negligible attenuation. The detection of TeV--PeV diffuse astrophysical neutrino emission by IceCube~\cite{Aartsen:2013jdh, IceCube:2014stg, Aartsen:2015knd, IceCube:2016umi, IceCube:2020acn, Abbasi:2020jmh, Abbasi:2021qfz} 
has opened new avenues for both
astrophysics and particle physics.
Nevertheless, the origin of these neutrinos remains largely unknown, prompting numerous studies to identify their sources~\cite{Abbasi:2010rd, Aartsen:2013uuv, Aartsen:2014cva, 
ANTARES:2015moa, Aartsen:2016oji, IceCube:2017der, Aartsen:2018ywr, IceCube:2019cia, IceCube:2019lzm, IceCube:2020svz, Zhou:2021rhl, Li:2022vsb, Chang:2022hqj, Lu:2025vmk, PugazhendhiAD:2025cjj}. While IceCube has reported evidence for neutrino emission from the directions of TXS~0506+056 and NGC~1068~\cite{IceCube:2018dnn, IceCube:2018cha, IceCube:2019cia, icecube2022evidence}, these sources together with other less significant candidates contribute only a small fraction to the total all-sky diffuse neutrino flux.
Recently, IceCube has detected neutrino emission from the Galactic Plane (GP) at a significance of $4.5\sigma$~\cite{IceCube:2023ame}. This GP signal is consistent with a diffuse neutrino flux from the Milky Way, though it could alternatively originate from a population of unresolved point sources. To date, no individual Galactic point sources\footnote{Given the limited angular resolution of neutrino detectors, the term ``point source'' in this work also includes various discrete small-scale extended sources, such as extended SNRs and PWNs.} have been definitively detected.

The Large High Altitude Air Shower Observatory (LHAASO) is a square-kilometer-scale observatory for cosmic-ray and gamma-ray detection~\cite{LHAASO:2019qtb}. Its primary scientific goal is to unravel the origin of Galactic cosmic rays. The observatory comprises two main detector arrays: the Water Cherenkov Detector Array (WCDA), sensitive to gamma rays from 0.1 to 30~TeV, and the Kilometer Squared Array (KM2A), covering the energy range from 25~TeV to $\gtrsim 1$~PeV. Leveraging its high altitude, large effective area, and hybrid detector design, LHAASO has opened a new window into ultra-high-energy gamma-ray astronomy~\cite{LHAASO:2021gok,LHAASO:2021cbz,LHAASO:2023kyg}. The LHAASO Collaboration has released the first catalog of gamma-ray sources (1LHAASO), containing 90 sources~\cite{LHAASO:2023rpg}, including 69 detected by WCDA and 75 by KM2A.

At very high energies, the Klein-Nishina effect suppresses inverse Compton (IC) emission from electrons, making hadronic processes potentially dominant. Sources in the LHAASO catalog are promising candidates for Galactic PeVatrons. However, gamma-ray detection at PeV energies alone does not constitute definitive proof of PeVatrons. For instance, although LHAASO-KM2A has detected gamma rays with energies up to 1.1~PeV from the Crab Nebula, the hadronic origin for the TeV gamma-ray emission from this source is considered unlikely~\cite{LHAASO:2021cbz}.

The detection of neutrino point sources is crucial for resolving the cosmic-ray origin puzzle. Neutrino emission is generally produced in hadronic processes via proton-proton ($pp$) or proton-gamma ($p\gamma$) interactions. Consequently, neutrino detection implies the presence of accelerated cosmic rays at the source, making high-energy neutrino observations a definitive means of identifying cosmic-ray sources. The previous IceCube measurements of GP neutrino emission relied on cascade events~\cite{IceCube:2023ame}, which suffer from poor angular resolution, rendering it difficult to distinguish between contributions from discrete sources and genuinely diffuse emission.

Given the overlapping energy ranges of LHAASO and IceCube, their observations are intrinsically connected. Several studies have investigated the neutrino emission from LHAASO sources and whether LHAASO catalog sources could contribute to the observed GP neutrino emission \cite{Huang2022_1,Huang:2021hjc,IceCube:2022heu,Shao2023,Fang:2023ffx,yan2024insights,marinos2025simulating,Luque2025JCAP}. For instance, Ref.~\cite{Fang:2023ffx} derived expected neutrino fluxes from the observed gamma-ray emission of these sources, suggesting that GP neutrino emission is likely dominated by diffuse emission and unresolved hadronic sources. 
Ref.~\cite{yan2024insights} compared
the total gamma-ray flux from LHAASO and HESS GP sources with the gamma-ray emission inferred from IceCube GP neutrino fluxes (using the $\pi^0$ template) and found that resolved sources could contribute up to approximately two-thirds of the total flux inferred from neutrino measurements. 
Ref.~\cite{marinos2025simulating} used GALPROP to model the diffuse neutrino emission from the Galactic Plane, finding that it contributes only $22\%$--$42\%$ of the IceCube GP flux,  implying a significant contribution from unresolved sources.
However, Ref.~\cite{Luque2025JCAP} suggested that unresolved sources contribute sub-dominantly.

Further investigation is required to disentangle the contributions of individual sources from intrinsic diffuse emission in the observed GP neutrino flux. In hadronic processes, neutrino production is accompanied by gamma-ray emission. In the absence of absorption, a high-energy gamma-ray source must also be a neutrino source, and vice versa. This makes the very-high-energy (VHE) and ultra-high-energy (UHE) sources detected by LHAASO promising candidates for neutrino source searches.

In this work, we utilize ten years of IceCube muon-track data \cite{IceCube:2021xar} to search for neutrino emission from sources in the LHAASO catalog. Our objectives are to identify neutrino signals associated with LHAASO sources, use the resulting neutrino flux upper limits to constrain hadronic gamma-ray emission, and determine the contribution of LHAASO sources to the total GP neutrino flux. We employ both single-source and stacking analyses to search for neutrino signals and derive 95\% confidence level upper limits on the neutrino flux. This study provides the first estimation of the contribution of VHE/UHE sources to the GP neutrino flux based on IceCube data analysis, helping to distinguish between diffuse and point-source contributions to the total GP neutrino emission.

\section{Neutrino data from IceCube} \label{sec:data}
The IceCube Observatory, located at the South Pole, has been operating for over a decade. The detector is buried in the Antarctic ice sheet at depths ranging from approximately 1,500 to 2,500 meters below the surface. It comprises 5,160 digital optical modules (DOMs) suspended on 86 vertical cables (``strings''). IceCube identifies neutrino interactions in the vicinity of the detector through Cherenkov light emitted by relativistic charged secondary particles propagating through the deep, ultra-clear glacial ice~\cite{IceCube:2006tjp}. This unique location provides natural shielding from background radiation at the Earth's surface, rendering IceCube the most sensitive detector for neutrinos in the TeV to PeV energy range.

The IceCube Collaboration has released muon-track data collected from 2008 to 2018~\cite{IceCube:2021xar}. This dataset is particularly well-suited for neutrino point-source searches and has been utilized in the ten-year time-integrated search for point-like neutrino sources~\cite{IceCube:2019cia}. The dataset comprises three components: (1) experimental data events, (2) instrument response functions, and (3) detector uptime.

The data events contain reconstructed information for all muon events satisfying the selection criteria, including reconstructed energy $E$, direction (right ascension $\alpha$ and declination $\delta$), directional uncertainty $\sigma$, and arrival time $t$. The events are divided into ten ``seasons'' (IC40, IC59, IC79, IC86-I, and IC86-II through IC86-VII), where the numerical values indicate the number of installed detector strings. Each season encompasses approximately one year of data. The instrument response varies between seasons, particularly during the detector construction phase.
The instrument response functions consist of effective areas and smearing matrices. The effective area $A_{\rm eff}(E_\nu, \delta)$ is a function of the true neutrino energy $E_\nu$ and declination $\delta$. The smearing matrices provide the probability density $P(E | E_\nu, \delta)$ of reconstructing a muon energy $E$ given a true neutrino energy $E_\nu$ and declination $\delta$. The detector uptime records provide the effective data-taking time $t_k$ for each season $k$.
In this work, we utilize these ten years of muon-track data to search for neutrino signals from the LHAASO sources.

\section{Analysis and results}\label{sec:ana}
We perform an unbinned likelihood analysis~\cite{Braun:2008bg} of the neutrino data to assess the significance of neutrino signals from the directions of LHAASO sources. The analysis follows the standard approach used in IceCube point-source searches, comparing the signal-plus-background hypothesis against the background-only hypothesis. The likelihood function is constructed as the product of probability densities for all events:
\begin{equation}\label{eq:simple_likelihood}
\mathcal{L}(n_s) = \prod_{i} \left[ \frac{n_s}{N} S_i + \left(1-\frac{n_s}{N}\right) B_i \right],
\end{equation}
where $n_s$ is the expected number of signal events, $N$ is the total number of events, and $S_i$ and $B_i$ are the signal and background probability density functions (PDFs) for event $i$, respectively. The analysis is time-integrated over the ten-year dataset. Details of the signal and background PDFs, as well as the full mathematical formulation, are provided in Appendix~\ref{likelyhood}.
The $B_i$ term encompasses backgrounds from atmospheric muons, atmospheric neutrinos, and isotropic astrophysical neutrinos. In principle, the GP neutrino emission component should also be included in the likelihood analysis. However, we find that incorporating this component has negligible impact on our results (see Appendix~\ref{app:gpdiffuse}), as GP neutrino emission has not yet been detected in muon data. Therefore, we neglect this component in our main analysis. Omitting it theoretically yields more conservative limits.
\subsection{Single-source analysis}\label{sec:single}
We first search for neutrino signals from individual sources in the LHAASO catalog. In the single-source analysis, each source is modeled with a power-law spectrum and we allow both the number of signal events $n_s$ and the spectral index $\Gamma$ to vary when searching for neutrino signals from each source in the LHAASO catalog. The source coordinates and spatial extensions are taken directly from the values reported in the catalog. The catalog sources are divided into WCDA and KM2A subsamples, containing 64 and 74 sources, respectively. Sources with Galactic latitude $|b| > 20^\circ$ and those classified as high-frequency peaked BL Lacertae (HBL) objects in Table~2 of Cao et al.~\cite{LHAASO:2023rpg} are excluded. Detailed information about the selected sources is provided in Appendix~\ref{app catalogs}.

For all 90 LHAASO sources, we find no significant neutrino excess above background; the maximum test statistic (TS) value is 8.5. For two degrees of freedom, this corresponds to a pre-trial significance of $2.2\sigma$. Accounting for the trial factor from searching 90 sources, the post-trial significance is only $0.6\sigma$, providing no evidence for significant neutrino emission.
We present 95\% confidence level (C.L.) upper limits on the neutrino fluxes from these LHAASO sources. For each source, the flux upper limit is obtained by lowering $\ln(\mathcal{L})$ by 2.71 from the best-fit value. Representative results are listed in Table~\ref{tab:wcda}; complete results are provided in Appendix~\ref{app catalogs}.
\subsection{Constraints on hadronic gamma-ray emission}
\label{sec:hadronic}
According to the relationship between neutrino and gamma-ray fluxes in hadronic processes, we can convert upper limits on neutrino fluxes into constraints on hadronic gamma-ray emission. For Galactic sources, neutrinos and hadronic gamma rays are primarily produced through $pp$ interactions:
\begin{gather}
p + p \rightarrow \pi^0, \pi^+, \pi^- + \dots, \nonumber \\
\pi^\pm \rightarrow e^\pm + \nu_e / \bar{\nu}_e + \nu_\mu / \bar{\nu}_\mu + \bar{\nu}_\mu / \nu_\mu, \\
\pi^0 \rightarrow \gamma\gamma. \nonumber
\end{gather}
In $\pi^0 \to \gamma\gamma$ decay, each gamma ray carries approximately half of the $\pi^0$ energy. In charged pion decay, the three resulting neutrinos each carry approximately one-quarter of the pion energy. Consequently, the energies satisfy $E_\gamma \simeq 2E_\nu$, and the differential fluxes are related by $F_\gamma(E_\gamma) = \frac{1}{2} F_{\nu_\mu+\bar{\nu}_\mu}(E_\nu)$. Using this relation, we derive upper limits on hadronic gamma-ray fluxes for the LHAASO catalog sources and compare them with LHAASO observations. For this analysis, we fix the spectral index of each source to the value reported in the catalog and scan over $n_s$ to obtain the 95\% confidence level upper limit on the neutrino flux.

Following the neutrino--gamma-ray connection in hadronic scenarios, we convert the neutrino flux upper limits into constraints on hadronic gamma-ray emission. The results are presented in Table~\ref{tab:wcda}, where the last column shows the ratio of the derived hadronic flux upper limit to the observed LHAASO flux. Ratios below 100\% indicate that hadronic emission cannot account for the entire observed flux, implying the definite existence of a leptonic component (primarily inverse Compton emission from electrons). Sources with ratios exceeding 100\% are not listed as they are not constrained by IceCube observations.

In total, seven WCDA sources are effectively constrained. The strongest constraints are obtained for the Geminga and Crab pulsars, with hadronic contributions limited to 15.5\% and 52.4\%, respectively. These low hadronic fractions are physically reasonable, as pulsar wind nebulae are generally expected to be dominated by leptonic emission processes. Beyond pulsars and TeV halos, two unidentified sources (J1843$-$0335u and J2020+3649u) also have their hadronic fractions constrained. Our results thus provide valuable information for identifying the counterparts of these sources.
\subsection{Stacking analysis}\label{sec:stacking}

Since the single-source analysis does not reveal any significant signals, to enhance the detection sensitivity, we also perform a stacking analysis. In our analysis, we find that for a spectrum with index $\lesssim-2.5$, the muon-track data used in this work are most sensitive to energies around $E_\nu \lesssim 30\ \mathrm{TeV}$,
which are within the energy range of WCDA. In contrast, KM2A detects gamma rays above $25\ \mathrm{TeV}$ and is most sensitive around $\sim 100\ \mathrm{TeV}$. Therefore, performing a stacking analysis using KM2A source parameters would face issues of spectral extrapolation. We note that, for sources detected by both KM2A and WCDA, KM2A generally gives a steeper spectral index than WCDA. For example, for 1LHAASO J0534+3533, the KM2A spectral index is $-4.89$, which actually lies in the exponentially decaying part of the spectrum. Directly extrapolating such KM2A spectra to below $25\ \mathrm{TeV}$ would lead to unreliable results. The LHAASO catalog also points out that the power-law spectral shape can adequately describe the WCDA components, while it is not suitable for about 1/3 KM2A components. For this reason, we primarily base our stacking analysis on sources detected by WCDA.
In this analysis, we further exclude sources likely to be of leptonic origin, specifically, those associated with TeV halos or pulsars (PSRs) in the catalog\footnote{In fact, including TeV halos and pulsars would yield stronger constraints, as sources like Geminga that exhibit low hadronic contributions (see Table~\ref{tab:wcda}) will also be included in the sample.}. Following this selection, the WCDA sample contains 57 sources.

\begin{table*}[!htbp]
\centering
\caption{WCDA sources with constrained hadronic gamma-ray emission.}
\begin{tabular}{p{2.2cm}ll p{1.5cm}p{1.5cm}p{1.5cm}p{1.5cm}}
\hline
\hline
Source Name & Assoc. & Type & $\Gamma$ & $\Phi_{\gamma,\rm 12TeV}^{\text{UL}}$ & $\Phi_{\gamma,\rm 12TeV}^{\text{obs}}$ & Ratio (\%) \\
\hline
J0634+1741u & Geminga        & PWN/TeV Halo & 1.65 & 0.24 & 1.55 & 15.5 \\
J0534+2200u & Crab           & PWN          & 2.69 & 2.66 & 5.07 & 52.4 \\
J0542+2311u & HAWC J0543+233 & TeV Halo     & 1.95 & 0.8 & 1.39 & 57.4 \\
J2018+3643u & MGRO J2019+37  & PWN          & 1.94 & 0.86 & 1.49 & 57.9 \\
J1843-0335u & HESS J1843-033 & UNID         & 2.58 & 1.49 & 2.52 & 59.2 \\
J0703+1405  & 2HWC J0700+143 & TeV Halo     & 1.98 & 1.14 & 1.46 & 78.5 \\
J2020+3649u & VER J2019+368  & UNID         & 1.78 & 0.39 & 0.47 & 83.2 \\ 
\hline
\hline
\end{tabular}
\begin{tablenotes}
\item The $\Phi_{\gamma,\rm 12TeV }^{\text{UL}}$ and $\Phi_{\gamma,\rm 12TeV}^{\text{obs}}$ represent the upper-limit and observed differential flux at 12 TeV respectively. Both are in units of $10^{-17} \ \mathrm{cm}^{-2} \ \mathrm{s}^{-1} \ \mathrm{GeV}^{-1}$. The ratio is given by $\Phi_{\gamma,\rm 12TeV }^{\text{UL}}/\Phi_{\gamma,\rm 12TeV}^{\text{obs}}$.
\end{tablenotes}
\label{tab:wcda}
\end{table*}

\begin{figure}[t]
\centering
\includegraphics[width=0.485\textwidth]{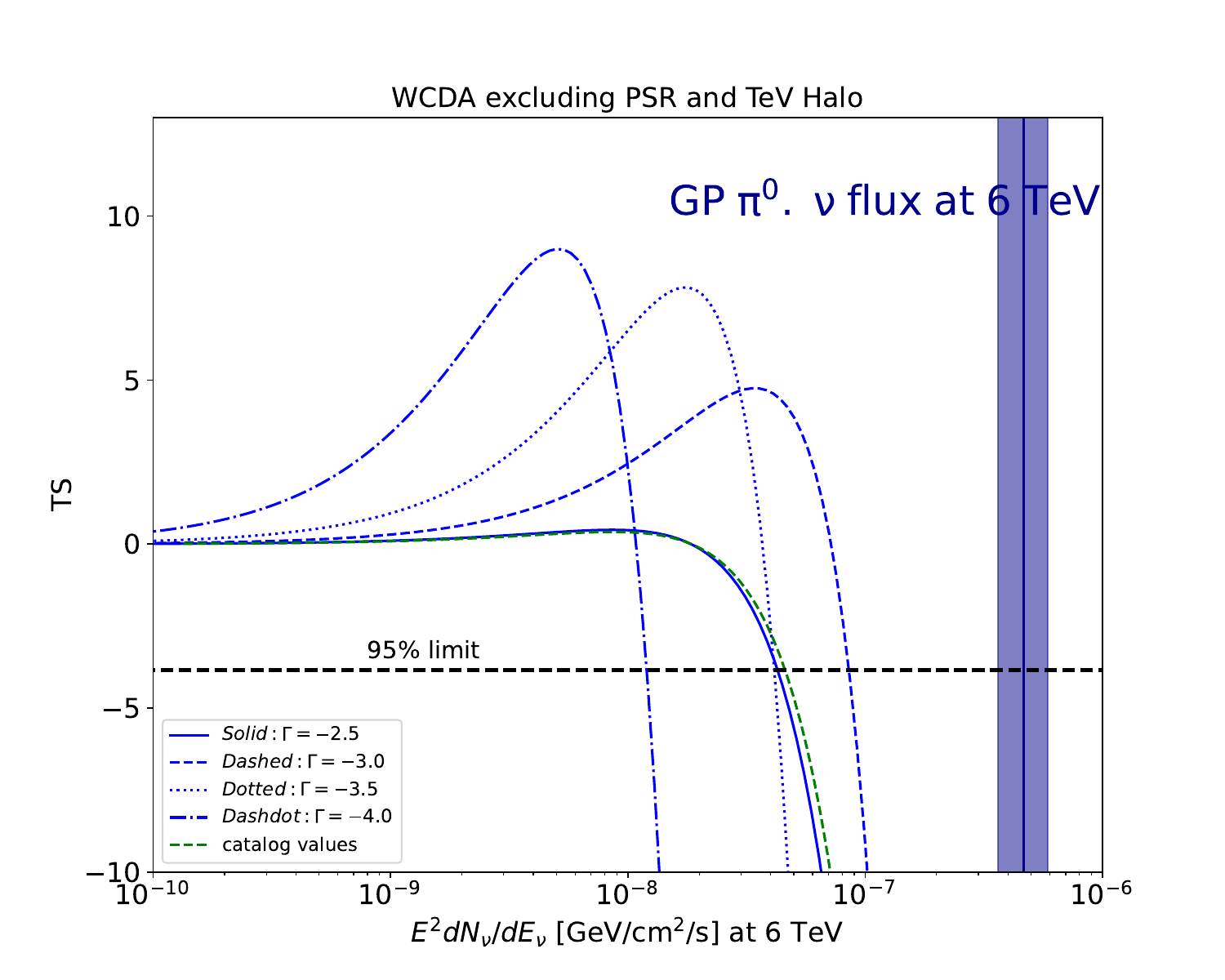}
\caption{The change of the TS value from the stacking analysis as a function of the total neutrino flux of the sources in the LHAASO WCDA catalog, for different neutrino spectral indices. The green dashed line represents the result using the actual detected spectral indices reported in the LHAASO catalog, while the blue lines are the results for different assumed average spectral indices. The diffuse Galactic plane neutrino flux measured by IceCube (using the $\pi^0$ model, vertical dark blue band) \cite{IceCube:2023ame} is also shown. The dashed line corresponds to ${\rm TS} = -3.84$, representing the 95\% confidence level TS threshold for the case of two degrees of freedom.}
\label{fig:tsflux}
\end{figure}

\begin{figure*}[t]
\centering
\includegraphics[width=0.6\textwidth]{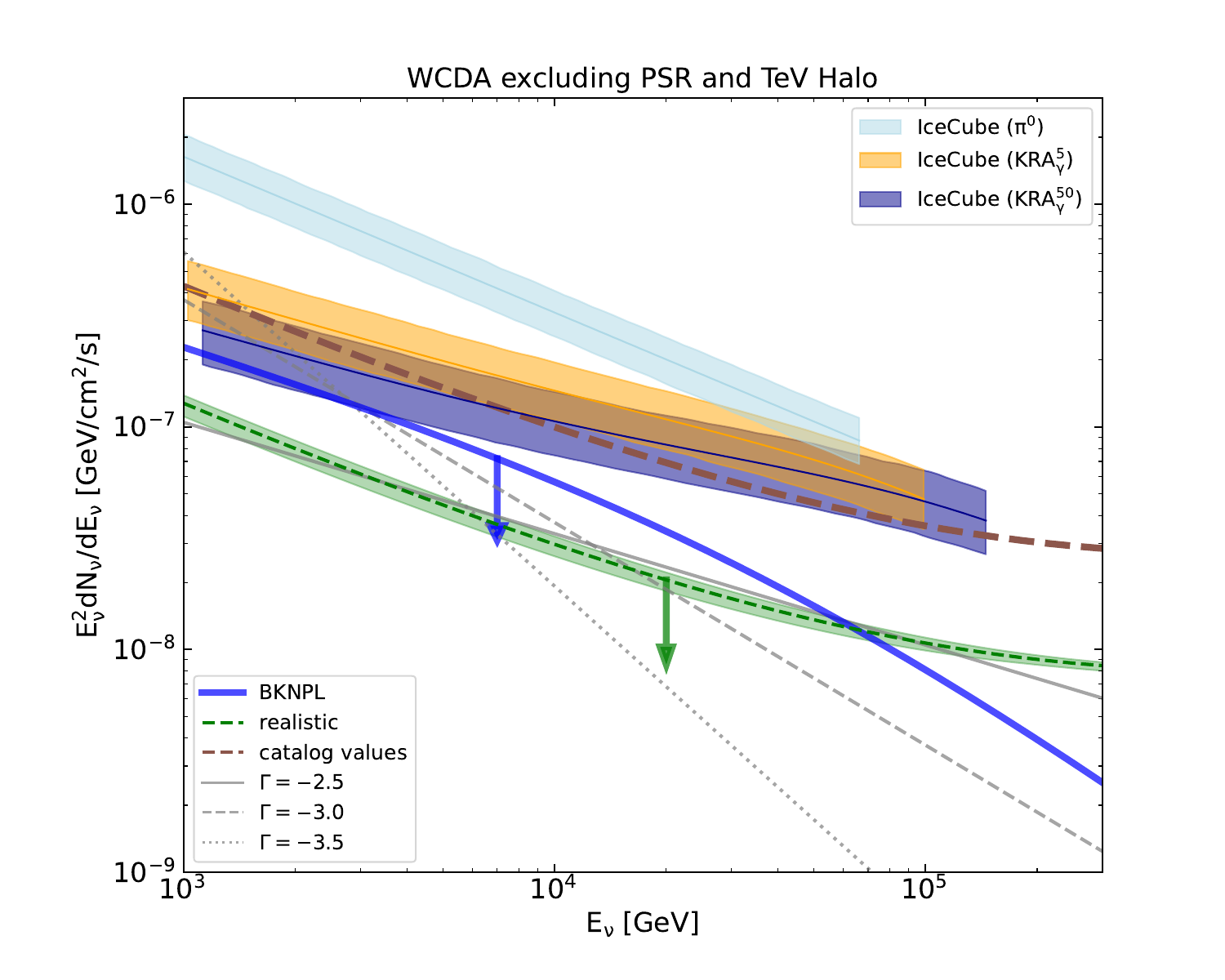}
\caption{The 95\% C.L. upper limits on the total neutrino flux from the sources in the LHAASO WCDA sample (green, blue and gray lines). The green dashed line corresponds to the result derived using the actual spectra measured by LHAASO, while the shaded bands represent the uncertainties from the LHAASO flux normalizations. 
The blue solid line corresponds to the upper limits obtained by assuming an average spectrum of a smoothly broken power-law with a break at 40~TeV. The gray lines assume power-law average spectra with different spectral indices.
These constraints are compared with the diffuse neutrino flux observed by IceCube from the Galactic plane \cite{IceCube:2023ame}, adopting the $\pi^0$ model (light blue), the $\text{KRA}^{5}_{\gamma}$ model (orange), and the $\text{KRA}^{50}_{\gamma}$ model (dark blue). 
Our results indicate the sources in the LHAASO catalog cannot account for all of the observed Galactic plane neutrino flux, contributing at most $\sim$20\% of the GP neutrinos compared to the $\pi^0$ result.}
\label{fig:wcda}
\end{figure*}

\begin{table}
\centering
\caption{Upper limits on the neutrino flux of WCDA sources.\label{tab:fraction}}
\begin{tabular}{cccc}
\hline
\hline
Spectrum & $\Phi_{6}^a$ & $R_{\pi^0}^b$ & $R_{\rm had}^c$ \\
& [$10^{-8}$ GeV cm$^{-2}$ s$^{-1}$] & (\%) & (\%) \\
\hline
$\Gamma = -2.5$ & 4.27 & 9.2 & 31.8 \\
$\Gamma = -3.0$ & 6.20 & 13.3 & 46.3 \\
$\Gamma = -3.5$ & 4.14 & 8.9 & 30.8 \\
Actual & 3.99 & 8.6 & 29.7 \\
BKNPL & 7.97 & 17.1 & 59.4 \\
\hline
\hline
\end{tabular}
\begin{tablenotes}
\footnotesize
\item $^a$The total neutrino flux ($\nu_\mu + \bar{\nu}_\mu$) at 6~TeV from all stacked WCDA sources (excluding pulsars and TeV halos).
\item $^b$The maximum allowed ratio of the total neutrino flux to the IceCube GP flux measured with the $\pi^0$ model at 6~TeV (i.e., $f_{\rm res}/f_{\pi^0}$).
\item $^c$The maximum allowed hadronic fraction of the total gamma-ray emission at 6~TeV, derived from the neutrino flux upper limits assuming the neutrino-to-gamma-ray connection in hadronic processes.
\end{tablenotes}
\end{table}

We adopt the differential flux at 6~TeV as the weighting factor for the signal PDF ($E_{\mathrm{ref}} = 6$~TeV). Tests with $E_{\mathrm{ref}} = 3$~TeV and 10~TeV show no significant impact on the main results. The stacking analysis results are presented in Figs.~\ref{fig:tsflux} and~\ref{fig:wcda}. In Fig.~\ref{fig:tsflux}, we scan over summed neutrino flux values $\Phi_{\nu,\rm tot} = \sum_j \Phi_{0,j}$ (at 6~TeV) to show how the test statistic ($TS = 2\ln[\mathcal{L}(\hat{n}_s)/\mathcal{L}(n_s=0)]$) varies with $\Phi_{\nu,\rm tot}$. A positive TS indicates that the model including a neutrino signal fits the observations better, suggesting a possible neutrino excess; while a negative TS means that the model with a neutrino signal added is less favored compared to the pure background model, implying that such a signal model parameter may be excluded. For one (two) additional degrees of freedom, the 95\% C.L. exclusion limit corresponds to $2\Delta\ln\mathcal{L} = -2.71$ ($-3.84$).

We note that our stacking analysis shows a mild positive TS, with a maximum of $TS \approx 10$, which might suggest a possible neutrino excess. However, the TS value remains relatively low. Moreover, the positive TS values occur preferentially for soft spectral indices (with the maximum TS occurring at $\Gamma = -4$), suggesting that this is most likely a spurious signal, even if not purely a statistical fluctuation. Such false positives can arise from imperfections in the energy-dependent background PDF. Our data-driven background estimation may deviate from the true distribution due to finite sampling resolution and statistical limitations. When the assumed spectrum is very soft, the energy signal PDF becomes increasingly similar to the background PDF (the conventional atmospheric neutrinos follow a power-law spectrum with $\gamma\approx3.7$), allowing the signal model to compensate for background imperfections and produce artificial excesses.

Fig.~\ref{fig:wcda} shows the 95\% C.L. upper limits on the total neutrino flux as a function of energy (green line). The shaded band represents the uncertainty arising from the normalization $N_0$ of the LHAASO spectra. To obtain this band, we randomly sample 100 sets of $N_0$ values within their error range and perform the stacking analysis using each realization to derive the weighting factors; the shaded region covers the full range of results from these 100 realizations.

The above results are obtained using the actual spectral indices measured by LHAASO-WCDA. To further evaluate how the results might vary under different spectral assumptions, we also use three different average spectral indices ($\Gamma = -2.5$, $-3.0$, and $-3.5$) to derive the upper limits. We assume the average spectral of WCDA sources to be these values, and in the stacking analysis, the spectral indices for all sources are set to be the same and equal to the chosen value. The results are shown as the gray lines in Fig.~\ref{fig:wcda}. These cases are used to assess the potential impact on the results due to the choice of spectral index. Please note that spectral indices as soft as $\Gamma = -3.5$ are not consistent with WCDA observations.

We note that the spectra above several tens of TeV measured by KM2A have steeper indices than those of the WCDA population. Directly extrapolating the WCDA power-law spectra to energies $>100$~TeV is therefore not appropriate. To consider this, we also employ a broken power-law (BKNPL) spectrum for the analysis. The spectral index below the break energy is set to $-2.5$ (approximately the index of the cumulative emission from WCDA sources), while the index above the break is set to $-3.25$ (approximately that of the cumulative emission from KM2A sources). The break energy is set to 40~TeV. As shown by the blue line in Fig.~\ref{fig:wcda}, the BKNPL spectrum yields a weaker constraint. The upper limits obtained under different spectral assumptions are summarized in Table~\ref{tab:fraction}.

\subsection{Contribution to the Galactic Plane neutrino flux}\label{sec:gp_contrib}

Although neutrino emission from the Galactic Plane has been detected by IceCube~\cite{IceCube:2023ame}, the limited angular resolution of cascade events precludes distinguishing whether the signal originates from truly diffuse processes (cosmic-ray interactions with the interstellar medium) or from a population of unresolved point sources. An important task is to disentangle the contributions from individual sources from that of genuinely diffuse emission. To this end, we compare our stacking analysis upper limits on the summed neutrino flux from LHAASO catalog sources with the diffuse neutrino emission measured by IceCube (shaded regions in Figs.~\ref{fig:wcda}). As shown in the figure, the upper limits on the total neutrino flux from LHAASO catalog sources in the 1--25~TeV range fall below the flux measured by IceCube.

Around $E_\nu \sim 10$~TeV, the LHAASO sources can account for at most approximately 20\% of the observed GP flux (compared with the $\pi^0$ model, which is more favored by the IceCube cascade data though the preference is not statistically significant). This indicates that the resolved point sources identified by LHAASO cannot explain the entirety of the GP neutrino emission. The observed emission must include a significant component from truly diffuse processes or a large population of unresolved point sources below the detection threshold (see also below for the discussion on the contribution from unresolved point sources).
Though such a conclusion has been claimed by previous works (e.g., \cite{Fang:2023ffx,yan2024insights}), our analysis of IceCube data provides a more direct and stronger constraint on the point-source contribution.
Especially, our upper limits lie below the GP neutrino flux derived using the KRA templates, being able to constrain the contribution of LHAASO sources even under these alternative scenarios.

In addition to showing the maximum fraction of the GP neutrino emission that can be contributed by the LHAASO-WCDA sources (second column), Table~\ref{tab:fraction} also presents the maximum fraction of the total cumulative gamma-ray emission from all WCDA sources that can be accounted for when converting the obtained neutrino flux upper limits back to gamma-rays (using the neutrino--gamma-ray connection shown in Sec.~\ref{sec:hadronic}). This effectively corresponds to the maximum hadronic fraction of these sources. Even for our weakest limit (the BKNPL result), hadronic emission can contribute at most approximately 60\% of the total gamma-ray emission from the WCDA sources; otherwise, the corresponding neutrino emission would exceed our derived upper limits. This suggests that the cumulative gamma-ray emission from all WCDA sources is likely dominated by leptonic processes (i.e., hadronic fraction is likely $<50$\%). Our analysis is based on the 10-year public data release, during which IceCube was fully constructed in only 7 years; the current dataset is already doubled in size. Analysis based on the latest data is promising to definitively determine whether the emission is lepton-dominated, unless a tentative signal is detected.

{To assess the total contribution from point sources, we must also consider unresolved sources below the LHAASO detection threshold. We model the spatial distribution and luminosity function of Galactic sources (see Appendix~\ref{app:unresolved}) to estimate the cumulative flux from such faint sources. Our calculation shows that the flux contribution from unresolved sources amounts to $\lesssim$40\% of that from resolved LHAASO sources (this value is also supported by more detailed dedicated analysis~\cite{Lipari2025}). Consequently, the total neutrino flux from all discrete sources (resolved plus unresolved) reaches at most $\sim$39\% of the observed GP neutrino flux ($f_{\rm res}/f_{\pi^0}\sim17.1\% \Rightarrow (f_{\rm res}+f_{\rm unres})/0.6f_{\pi^0} \sim 2.3f_{\rm res}/f_{\pi^0} \sim 39.3\% $, where the factor of 0.6 considers the fact that LHAASO only observes $\sim60\%$ of the GP), which is still far below the level required to explain the IceCube measurement. 
This estimate assumes that unresolved sources have the same hadronic fraction as resolved sources. If the unresolved sources have higher hadronic fraction, the conclusion here will be weakened. However, there is currently no observational or theoretical support for the higher hadronic fraction of the unresolved source population.
Therefore, our result strongly supports that the bulk of the GP neutrino flux originates mainly from truly diffuse process, i.e., cosmic-ray interactions with the interstellar medium, rather than from point sources.}

\subsection{KM2A source analysis}\label{sec:km2a}

\begin{figure}[t]
\centering
\includegraphics[width=0.485\textwidth]{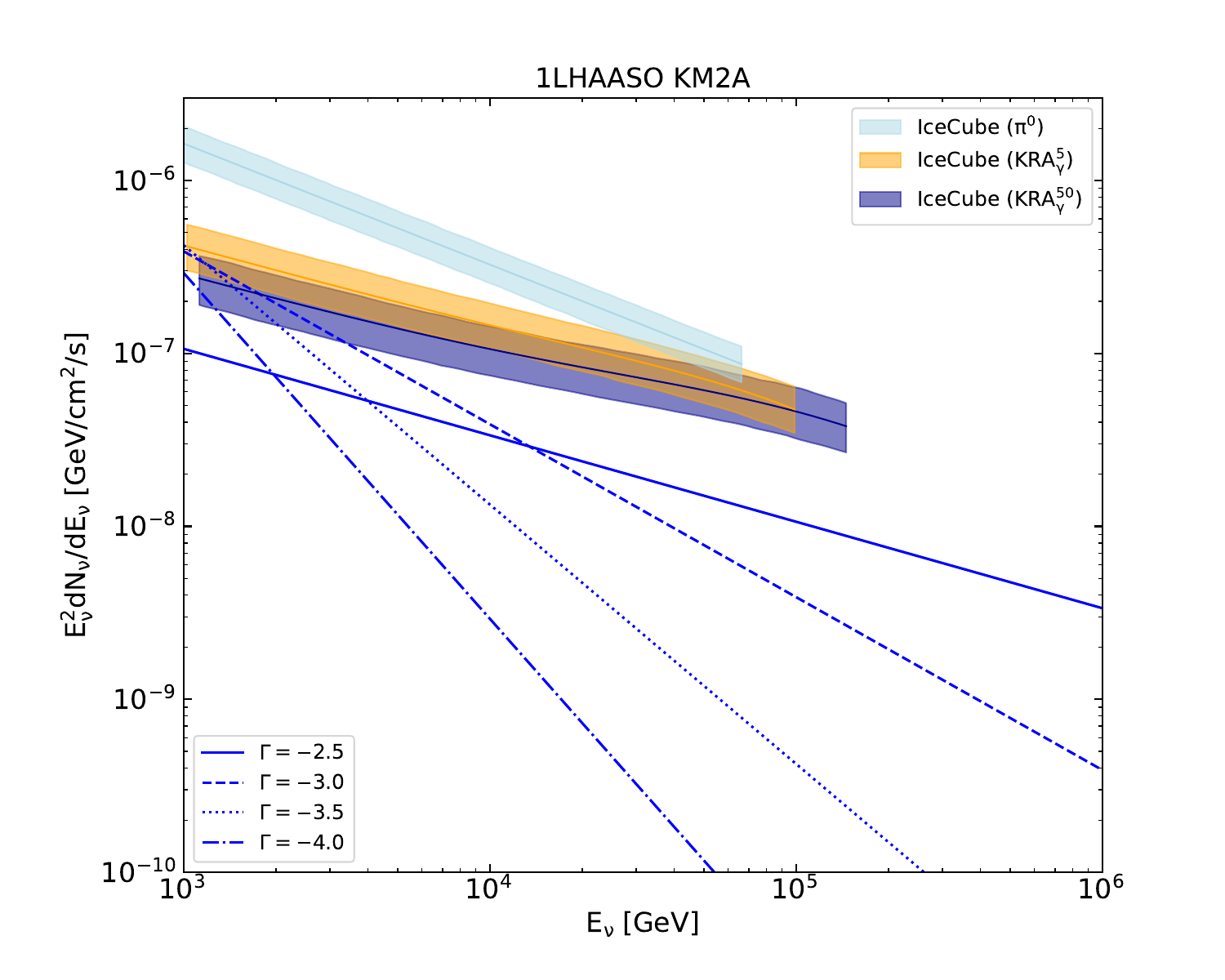}
\caption{Similar to Fig.~\ref{fig:wcda}, but for the LHAASO KM2A sources.}
\label{fig:km2a}
\end{figure}

As discussed above, the primary sensitivity of the IceCube muon-track data lies in the $\lesssim$30~TeV range, which does not overlap well with the KM2A energy coverage ($>25$~TeV). Consequently, a stacking analysis using KM2A source spectra would require significant extrapolation. Nevertheless, as a supplementary analysis, we also present constraints for KM2A sources using assumed average spectra.
We adopt a reference energy of $E_{\mathrm{ref}} = 100$~TeV for the weighting factor in the signal PDF. Fig.~\ref{fig:km2a} shows the constraints for KM2A sources under different spectral assumption. It can be seen that the analysis of KM2A sources yields results consistent with or better than those based on WCDA sources, supporting the conclusion that very-high-energy gamma-ray sources cannot account for the entirety of the GP neutrino flux.

In the current 1LHAASO catalog, the energy spectra for WCDA and KM2A are reported separately, each described by a power-law function. In future updated versions of the LHAASO catalog, a unified spectral model spanning the full energy range of both arrays would effectively resolve the spectral extrapolation issue encountered in this analysis.

\section{Summary and conclusions}\label{sec:discuss}
As strong candidates for Galactic PeVatrons, LHAASO sources may emit high-energy neutrinos through hadronic processes and are promising neutrino sources. In this work, we use 10 years of IceCube muon-track data to search for neutrino emission from the sources in the LHAASO catalog. LHAASO is one of the most sensitive VHE/UHE observatories in the world and its field of view largely overlaps with the most sensitive sky region of the 10-year IceCube data, providing an ideal source sample for studying the contribution of point sources to the Galactic plane neutrino flux. 
We analyze the IceCube data in the directions of these LHAASO sources using both single-source and stacking likelihood analyses. We find that neither analysis reveals significant neutrino emission signals. Accordingly, we present the 95\% C.L. upper limit on the neutrino flux for each source. Using the multi-messenger connection between neutrino flux and gamma-ray flux in hadronic models, we further convert the neutrino flux upper limits into constraints on the hadronic gamma-ray emission from these sources. 

Our conclusions include: (1) For seven WCDA sources, the hadronic contribution to the $\gamma$-ray emission is constrained to less than $100\%$ (Table~\ref{tab:wcda}), requiring a leptonic component; for the two unidentified sources among them, this provides valuable clues to their classification.
(2) The total hadronic contribution to the cumulative $\gamma$-ray emission from all WCDA sources is constrained to at most $\sim60\%$ (Table~\ref{tab:fraction}), suggesting that the bulk of their gamma-ray emission is likely lepton-dominated, or at least that leptonic and hadronic contributions may be comparable.
(3) The LHAASO WCDA sources contribute $\lesssim20\%$ to the diffuse Galactic Plane neutrino flux (at $E_\nu \sim 6$~TeV), significantly below the level required to explain the IceCube measurement.
(4) Unresolved sources below the detection threshold contribute an estimated $\lesssim40\%$ of the flux from resolved sources; accounting for these, the total neutrino flux from all discrete sources (resolved plus unresolved) reaches at most $\sim39\%$ of the observed GP neutrino flux, indicating that the bulk of the GP neutrino emission primarily originates from cosmic-ray interactions with the interstellar medium, rather than from discrete sources.

While individual Galactic neutrino point sources have yet to be detected, our work demonstrates that current IceCube observations already provide valuable insights into discriminating between discrete and diffuse emission components, thereby shedding light on the origin of the Galactic Plane neutrino emission.

\begin{acknowledgments}
We would like to thank Bei Zhou, Tian-Qi Huang and Ben-Yang Zhu for the helpful discussions.
This work is supported by the National Key Research and Development Program of China (Grant No. 2022YFF0503304) and Innovation
Project of Guangxi Graduate Education (YCSW2025133).
\end{acknowledgments}

\bibliography{sample701}{}
\bibliographystyle{aasjournalv7}
\clearpage
\begin{appendix}

\section{Maximum likelihood analysis}\label{likelyhood}
The likelihood function is given by the product of probability density of each muon-track event (indexed by $i$) in the ten data seasons (indexed by $k$):
\begin{equation}\label{eq1}
    \mathcal{L}(n_{s}) = \prod_{k} \prod_{i \in k} 
    \left[\frac{n_s^k}{N_k} S_i^k + \left(1 - \frac{n_s^k}{N_k}\right) B_i^k \right]
\end{equation}
where $n_s^k$ and $N_k$ denote the expected number of signal neutrinos and the total number of neutrino events in sample $k$, respectively. 
The number of signal neutrinos is related to the spectral model $\Phi(E_\nu)$ by $n_{s}^{k} = t_{k}\times \int A_{\rm eff}^{k} \left ( E_{\nu },\delta _{s}  \right ) \Phi \left ( E_{\nu } \right ) dE_{\nu}$. For a power-law spectral model, the neutrino flux is parameterized as $\Phi\left ( E_{\nu }  \right ) =\Phi _{0}\times \left ( E_{\nu } / E_{\rm ref}  \right )^{\Gamma }$ and $\Phi _{0}$ is the differential flux at reference energy.

The $S_i^k$ and $B_i^k$ in Eq.~(\ref{eq1}) are the signal and background probability density functions (PDFs) of event $i$ in the season $k$, representing the probability density of measuring the event for given source and background models. Both the signal and background PDFs can be factorized into two parts:
\begin{align}\label{eq2}
    S_i &= S^{\rm spat}\left ( \vec{x}_i\mid\sigma_i,\vec{x}_s,\sigma_s\right ) \times S^{\rm ener}\left ( E_i\mid \vec{x}_s,\Gamma   \right )   \\
    B_{i} &=B^{\rm spat} \left ( \delta _{i}  \right ) \times B^{\rm ener} \left ( E_{i} \mid \delta _{i}  \right )
\end{align}
The spatial term of the signal PDF $S_{ij}^{\rm spat}$ depends on the angular uncertainty of the event $\sigma_{i}$ and the angular difference $\phi$ between the event's reconstructed direction $\vec{x}_i$ and the direction of the source $\vec{x}_j$. We may consider the contributions of multiple sources to the signal PDF, especially in the stacking analysis below, we use the index $j$ to represent the $j$-th source. Usually, a two-dimensional Gaussian form is adopted for the spatial signal PDF. After considering the source's spatial extension $\sigma _{j}$, it is given by~\cite{Huang:2021hjc}
\begin{equation}\label{eq4}
    S_{ij}^{\rm spat}\left ( \vec{x}_i\mid\sigma_i,\vec{x}_j,\sigma_j\right )=\frac{1}{2\pi \left ( \sigma _{i}^{2} +  \sigma _{j}^{2}  \right ) } \text{exp}\left [ -\frac{\phi ^{2} }{2\left ( \sigma _{i}^{2}+  \sigma _{j}^{2}    \right ) }  \right ]\frac{\phi }{\sin \phi }
\end{equation}
The energy term $S_{ij}^{\rm ener}$ describes the PDF of measuring a event with a reconstructed energy $E_i$ for a source at the direction $\vec{x}_j$ with a given energy spectrum. Assuming a power law spectrum of spectral index $\Gamma$, it is given by~\cite{Huang:2021hjc}
\begin{equation}\label{eq5}
    S_{ij}^{\rm ener}\left ( E_{i}\mid \vec{x}_{j}, \Gamma \right )=\frac{\int E_{\nu }^{-\Gamma }A_{\rm eff}^{k}\left ( E_{\nu },\delta _j \right )dE_{\nu }\cdot P_{k} \left ( E_{i}\mid E_{\nu },\delta _{j}    \right )   }{\int E_{\nu }^{-\Gamma }A_{\rm eff}^{k}\left ( E_{\nu },\delta _{j}   \right )dE_{\nu }}
\end{equation}

The background PDF is directly derived from the experimental data. This is valid because only a very small number of neutrino sources have been detected so far, and the IceCube data are almost entirely dominated by background events.
Given that IceCube is located at the South pole, background events are uniformly distributed in right ascension and the spatial background PDF depends only on declination.
The spatial and energy terms are given by 
\begin{align}\label{eq6}
    B_i^{\rm spat}\left (\delta_i\right) &=\frac{1}{2\pi }\frac{N^k_{i}}{N_{k}\times \Delta \sin{\delta_{i}  }  } \\
    B_i^{\rm ener}\left ( E_i|\delta_i  \right ) &=\frac{N_{ij}^{k}  }{N^k_{i} \times \Delta \ln{E}_j }
\end{align}
{where $N^k_{i}$ is the number of events in declination band $i$, $N_{ij}^k$ is the number of events within declination band $i$ and reconstructed energy bin $j$ for sample $k$ and $\Delta \ln{E}_j$ is the corresponding energy bin width. Here, we adopt the same background treatment as the SkyLLH software package~\cite{IceCube:2023ihk}. The declination bins are chosen to be slightly finer near the celestial equator. Similar to the energy bins, their widths depend on the specific event sample. Finally, both the declination and energy distributions are smoothed, using spline interpolation and Gaussian convolution, respectively.}
We have checked part of our results using the SkyLLH package and obtained consistent results~\cite{IceCube:2023ihk}.

For stacking analysis of multiple sources, the signal PDF becomes the weighted sum of the contributions from all sources~\cite{Abbasi:2010rd,Aartsen:2013uuv}
\begin{equation}\label{eq7}
S_{i}^{k}=\frac{\sum _{j}\omega _{j,\text{model}}\omega _{j,\text{acc}}^{k}S_{ij}^{k}}{\sum _{j}\omega _{j,\text{model}}\omega _{j,\text{acc}}^{k}}, 
\end{equation}
where $\omega _{j,\text{model}}$ and $\omega _{j,\text{acc}}^{k}$ are weighting factors. The
$\omega_{j,\text{acc}}^{k} = t_{k} \times \int A_{\rm eff}^{k}(E_{\nu},\delta_{j})(E_{\nu }/E_{\rm ref})^{\Gamma_j}dE_{\nu }$ reflects the detector’s acceptance.
The $\omega_{j,\text{model}}$ describes the expected relative signal intensities between sources at the reference energy $E_{\rm ref}$, which is determined by the intrinsic properties of the sources. 
In this paper, we examine the hadronic contribution to the LHAASO sources by assuming that all their gamma-ray emissions originate from hadronic processes. 
Therefore, we assume that the high-energy neutrino flux is proportional to the LHAASO gamma-ray flux $f_{\rm LHAASO}$ and we have $\omega _{j,\text{model}}= f_{\rm LHAASO}$. Finally, the test statistic (TS) of our analysis is $TS=2\log \left[{\mathcal{L}\left( \hat{n}_{s}\right)}/{\mathcal{L}\left( n_{s}=0\right)}\right]$, where the denominator is the background or null hypothesis that all the events come from background, and the $\hat{n}_{s}$ is the best-fit number of signal events.

\section{Simultaneously fit the GP point-source and diffuse neutrino components}\label{app:gpdiffuse}
In addition to the potential neutrino point sources we are seeking, there also exists diffuse neutrino emission in the Galactic plane region. Our results in the main text tend to suggest that the GP neutrino flux may be dominated by the diffuse component. Therefore, when modeling the total neutrino emission from the GP, it is better to include contributions from both point sources and diffuse emission.
Since the diffuse neutrino emission from the GP has not yet been detected in the IceCube muon-track data, for simplicity, we have neglected its contribution in the main text. This generally leads to more conservative results for our conclusions. However, in order to obtain a more self-consistent result, in this appendix we attempt to include the LHAASO point sources, the diffuse neutrino emission from the GP, and the background component together in the model for the likelihood analysis.

In typical point-source analyses, it is usually assumed that point sources contribute negligibly to the total number of events within a given declination band, and therefore their effect can be ignored in background estimation. However, for the GP diffuse signal, since its emission extends over a large region, the cumulative flux contribution may not be negligible. Thus, the signal contamination to the background must be taken into account \cite{Pinat:2017wxs,IceCube:2017trr}. The true background $B_i$ should be the data-driven background $\tilde{D}_i$ with the signal “contamination” term $\tilde{S}_i$ subtracted:
$$
\tilde{D}_{i}=\frac{n_{s}}{N}\tilde{S}_{i}+\left (1-\frac{n_{s} }{N}\right )B_{i}
$$
Since our model includes both point-source and diffuse signal components, the background PDF directly obtained from the observational data (i.e., $\tilde{D}$) is actually a mixture of three parts:
$$
\tilde{D} =\frac{n_{s1} }{N}\tilde{S}_{1}+\left ( 1-\frac{n_{s1} }{N}-\frac{n_{s2} }{N} \right )B+\frac{n_{s2} }{N}\tilde{S}_{2}
$$
Here, $\tilde{S}_{1}$ and $\tilde{S}_{2}$ represent the signal PDFs of the point-source and diffuse components, respectively, after scrambling in the declination direction.
By solving for the background $B$ from the above equation and substituting it into the total likelihood function, we obtain the likelihood function for the “two-signal-component” case:
$$
L\left ( n_{s1},n_{s2},\gamma \right )=\prod_{i=1}^{N} \left [ \frac{n_{s1} }{N}S_{1}+  \frac{n_{s2} }{N}S_{2} + \tilde{D} - \left( \frac{n_{s1}}{N}\tilde{S}_{1} + \frac{n_{s2}}{N}\tilde{S}_{2} \right) \right ]
$$

\begin{figure}[htbp]
\centering
\begin{minipage}[t]{0.49\textwidth}
  \centering
  \includegraphics[width=\textwidth]{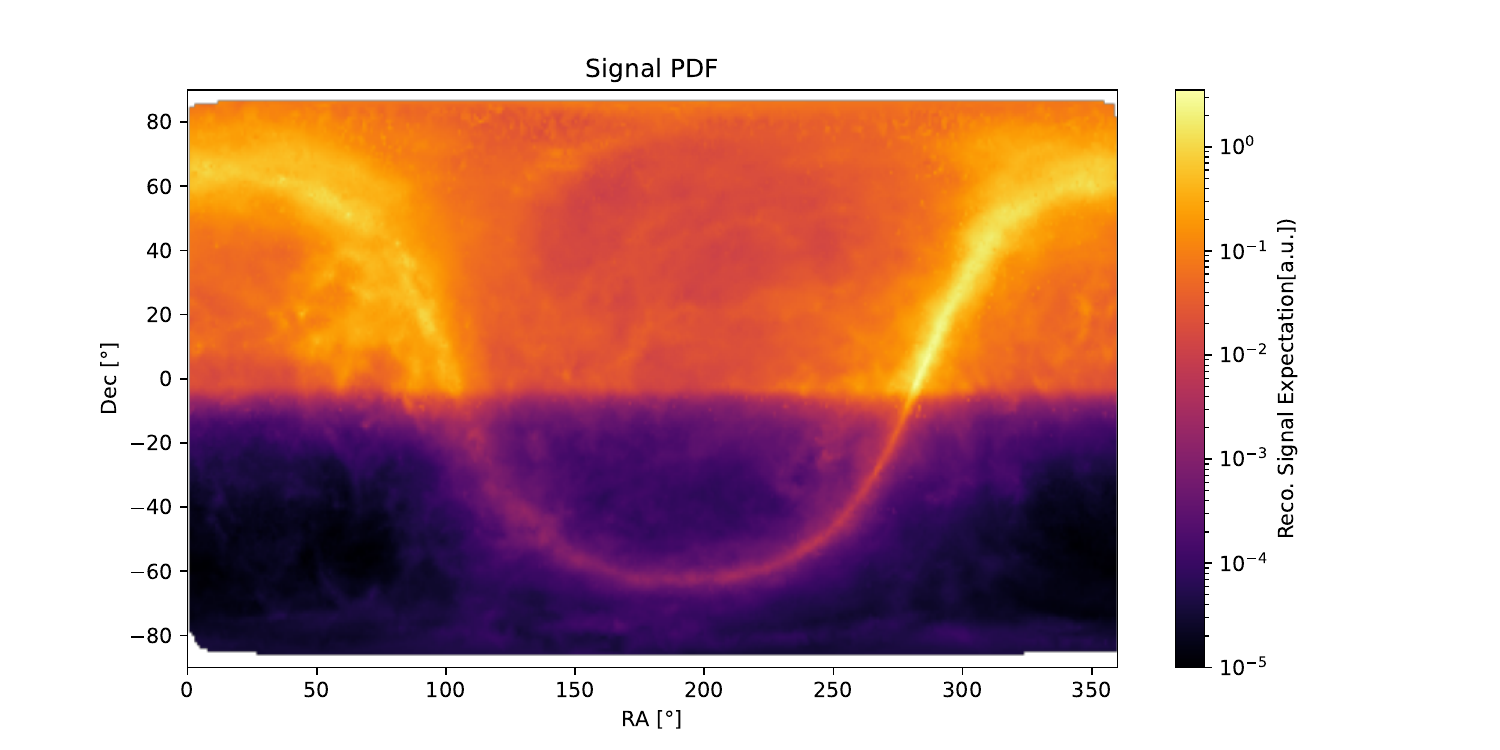}
  \caption{The signal PDF for the $\pi^0$ map after convolution.}
  \label{fig:si}
\end{minipage}
\hfill
\begin{minipage}[t]{0.49\textwidth}
  \centering
  \includegraphics[width=\textwidth]{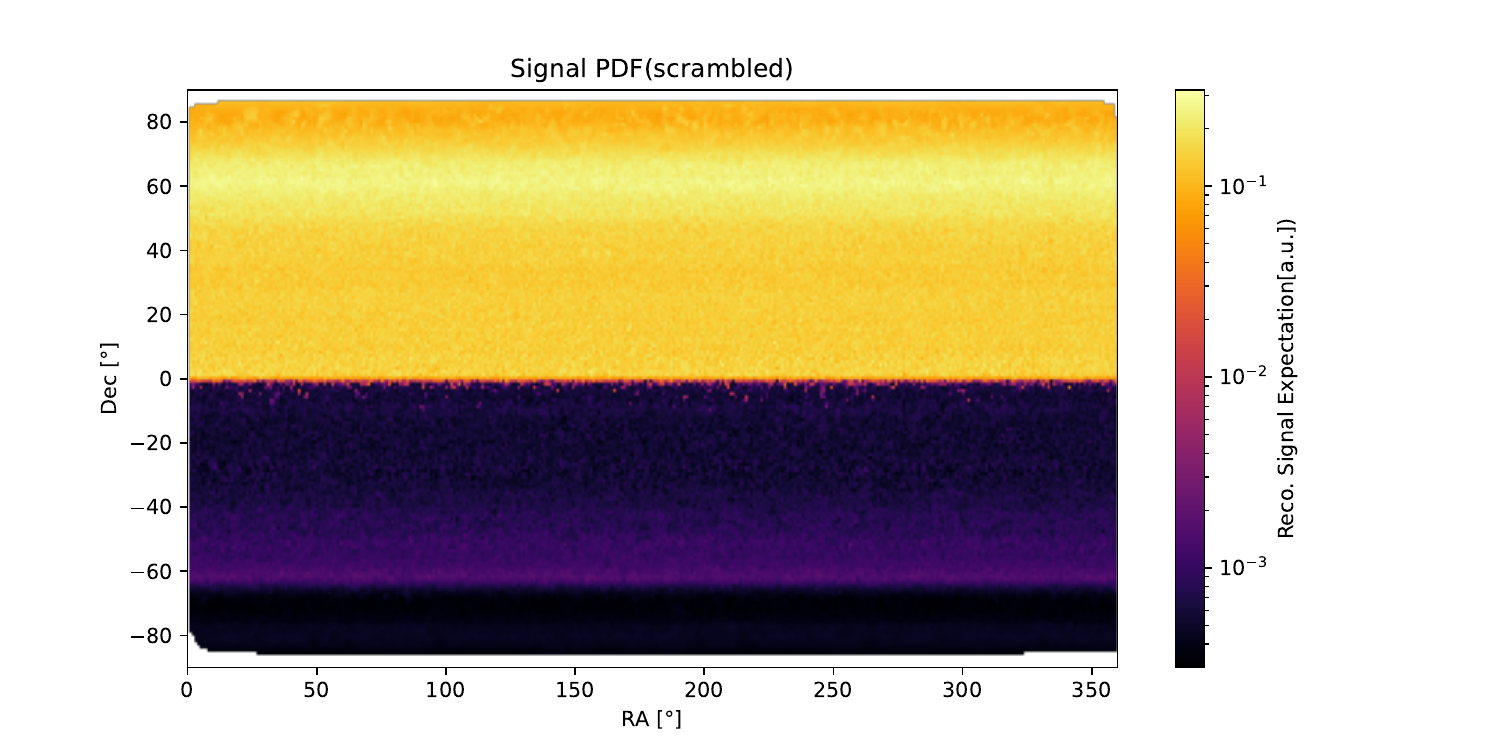}
  \caption{The signal PDF after scrambling.}
  \label{fig:si_tilde}
\end{minipage}
\end{figure}

For the diffuse component term $\tilde{S}_{2}$, we assume that the GP neutrino flux follows a single power-law spectrum with a spectral index of $-2.7$ across the entire sky, and its spatial distribution is modeled using the $\pi^0$ decay template. In the processing procedure, the $\pi^0$ decay template is first multiplied by the detector’s effective area  for each spatial pixel. Then, using each event’s directional uncertainty $\sigma_i$, the resulting all-sky map is convolved with the point spread function (PSF). Finally, the convolved all-sky map is averaged over right ascension to obtain the scrambled sky map (i.e., $\tilde{S}_2$). As shown in Fig.~\ref{fig:si} and Fig.~\ref{fig:si_tilde}, after averaging over right ascension, the spatial structural features of the $\pi^0$ decay template along right ascension are smoothed out.

We first performed a fit using only the $\pi^0$ decay component (without including the LHAASO point-source component) and found that the results are consistent with Ref.~\cite{Li:2025ank} (we obtained $\hat{n}_s = 637, \; \sigma_{\rm pre} = 1.56$). This demonstrates that our analysis procedure for the diffuse component (such as the scrambled PDF and the convolution of the $\pi^0$ decay map) is reliable.

In the analysis that includes both point-source and diffuse components simultaneously, when scanning over the number of point-source events $n_{s1}$, we treat the number of diffuse-template events $n_{s2}$ as a free parameter and maximize the likelihood function. We mainly tested the case of a BKNPL average spectrum and the case using the actual spectra measured by LHAASO. We find that after including the diffuse emission component, the constraints obtained for the LHAASO point-source component are almost unaffected. Therefore, the results presented in the main text where the diffuse component is neglected are robust.

\section{Calculation of the total flux from unresolved point sources}
\label{app:unresolved}

In this appendix, we describe the details for calculating the cumulative flux from unresolved point sources in the Galactic plane. The calculation involves modeling the spatial distribution and luminosity function of Galactic sources, then deriving both the number of detectable/unresolved sources and their flux contribution.

\subsection{Spatial distribution and luminosity function of sources}

We assume that the sources are distributed in the Galactic disk with a radial density profile following the pulsar distribution observed in \cite{Lorimer2006MNRAS}:
\begin{equation}
\label{eq:rho_R}
\rho(r') = A \left(\frac{r'}{R_\odot}\right)^B \exp\left[-C\left(\frac{r'-R_\odot}{R_\odot}\right)\right] \,,
\end{equation}
where $R_\odot = 8.5$~kpc is the Galactocentric radius of the Sun, $A = 41 \pm 5$~kpc$^{-2}$, $B = 1.9 \pm 0.3$, and $C = 5.0 \pm 0.6$ \cite{Lorimer2006MNRAS}. The sources are distributed vertically with an exponential profile $\propto \exp(-|z|/H)$ and a scale height $H = 0.2$~kpc \cite{Cataldo2020ApJ,Vecchiotti2022}.

The source number density is $n(r',z) = n_0(r') \exp(-|z|/H)$, where $n_0(r')$ is the midplane density. The primed quantity indicates that we are working in Galactocentric coordinates $(r', \phi', z)$, with $(r', \phi')$ the planar coordinates and $z$ the vertical coordinate. The Earth is located at $z=0$. {The surface probability density $f(r')=2H n_0(r')/N$ is normalized such that:
\begin{equation}
\int_{-\infty}^{+\infty} \int_0^{2\pi} \int_0^{R_{\rm max}} \frac{f(r')}{2H} \exp(-|z|/H) \, r' \, dr' \, d\phi' \, dz = 1 \,,
\end{equation}
where $R_{\rm max} = 30$~kpc is the maximum disk radius considered.}

We adopt an exponentially truncated power-law for the source luminosity function \cite{Lipari2025}:
\begin{equation}
\label{eq:luminosity_func}
g(L) = D \cdot L^{-\alpha} \exp(-L/L_{\rm cut}) \,, \quad L \geq L_{\rm min} \,,
\end{equation}
where $\alpha$ is the power-law index, $L_{\rm cut}$ is the cutoff luminosity, and $L_{\rm min}$ is the minimum luminosity.
In this work we adopt $L_{\rm min}=10^{29}\,{\rm erg/s}$, however, a lower value of $L_{\rm min}$ would not affect the integrated flux too much provided that $\alpha<2.0$.
The normalization $D$ guarantees $\int_{L_{\rm min}}^{\infty} g(L) \, dL = 1$ and can be calculated analytically:
\begin{equation}
D = \frac{1}{L_{\rm cut}^{1-\alpha} \, \Gamma(1-\alpha, x_{\rm min})} \,,
\end{equation}
where $x_{\rm min} = L_{\rm min}/L_{\rm cut}$ and $\Gamma(s, x)$ is the upper incomplete gamma function.

\subsection{Number of detectable sources and unresolved flux contribution}

For a source at distance $s$ from the Earth, the detected flux is $\Phi = L/(4\pi s^2 \langle E \rangle)$, where $\langle E \rangle$ is the average photon energy. For the LHAASO energy range of $1-25$~TeV assuming a spectral index of $\gamma = 2.5$, the mean energy is $\langle E \rangle \simeq 2.42$~TeV $\approx 3.88$~erg. For a detector with flux threshold $\Phi_{\rm th}$, the corresponding luminosity threshold is:
\begin{equation}
\label{eq:L_th}
L_{\rm th}(s) = 4\pi s^2 \Phi_{\rm th} \langle E \rangle \,.
\end{equation}

The number of sources detectable above flux threshold $\Phi_{\rm th}$ is obtained by integrating over all volumes where $L > L_{\rm th}(s)$:
\begin{equation}
\label{eq:N_det}
N_{\rm det}(\Phi_{\rm th}) = N \int_{-\infty}^{+\infty} \int_0^{2\pi} \int_0^{R_{\rm max}} \frac{f(r')}{2H} \exp(-|z|/H) \, \mathcal{G}(L_{\rm th}(s)) \, r' \, dr' \, d\phi' \, dz \,.
\end{equation}
where $\mathcal{G}(L_{\rm th}) = \int_{L_{\rm th}}^{\infty} g(L) \, dL$ is the fraction of sources at distance $s$ with luminosity above the threshold $L_{\rm th}$.
Note that $L_{\rm th}$ depends on $s$ (which depends on $r'$, $\phi'$, and $z$), so the integral over luminosity cannot be factored out.

Similarly, the total flux from detectable sources above threshold $\Phi_{\rm th}$ in the entire Galaxy is:
\begin{equation}
F_{\rm res}(\Phi_{\rm th}) = \frac{N}{4\pi \langle E \rangle} \int_{-\infty}^{+\infty} \int_{\phi_{\rm min}}^{\phi_{\rm max}} \int_{r_{\rm min}}^{r_{\rm max}} \frac{f(r'(r,\phi)) \, \exp(-|z|/H)}{2H} \frac{G(L_{\rm th}(s))}{s^2} \, r \, dr \, d\phi \, dz \,,
\end{equation}
where $G(L_{\rm th}) = \int_{L_{\rm th}}^{\infty} L \, g(L) \, dL$ is the cumulative luminosity of sources with luminosity above $L_{\rm th}$ and  $s = \sqrt{r^2 + z^2}$ is the line of sight distance. The angular range $[\phi_{\rm min}, \phi_{\rm max}] = [10^\circ, 235^\circ]$ in the above equation accounts for the LHAASO observational field of view. Here, for computational convenience, we have adopted the heliocentric coordinate system $(r, \phi, z)$. To calculate the flux from {unresolved} sources, one simply needs to change the integration range of the luminosity integral $G(L_{\rm th})$ in the above equation to from $L_{\rm min}$ to $L_{\rm th}(s)$.

\begin{figure}[ht]
\centering
\includegraphics[width=0.5\textwidth]{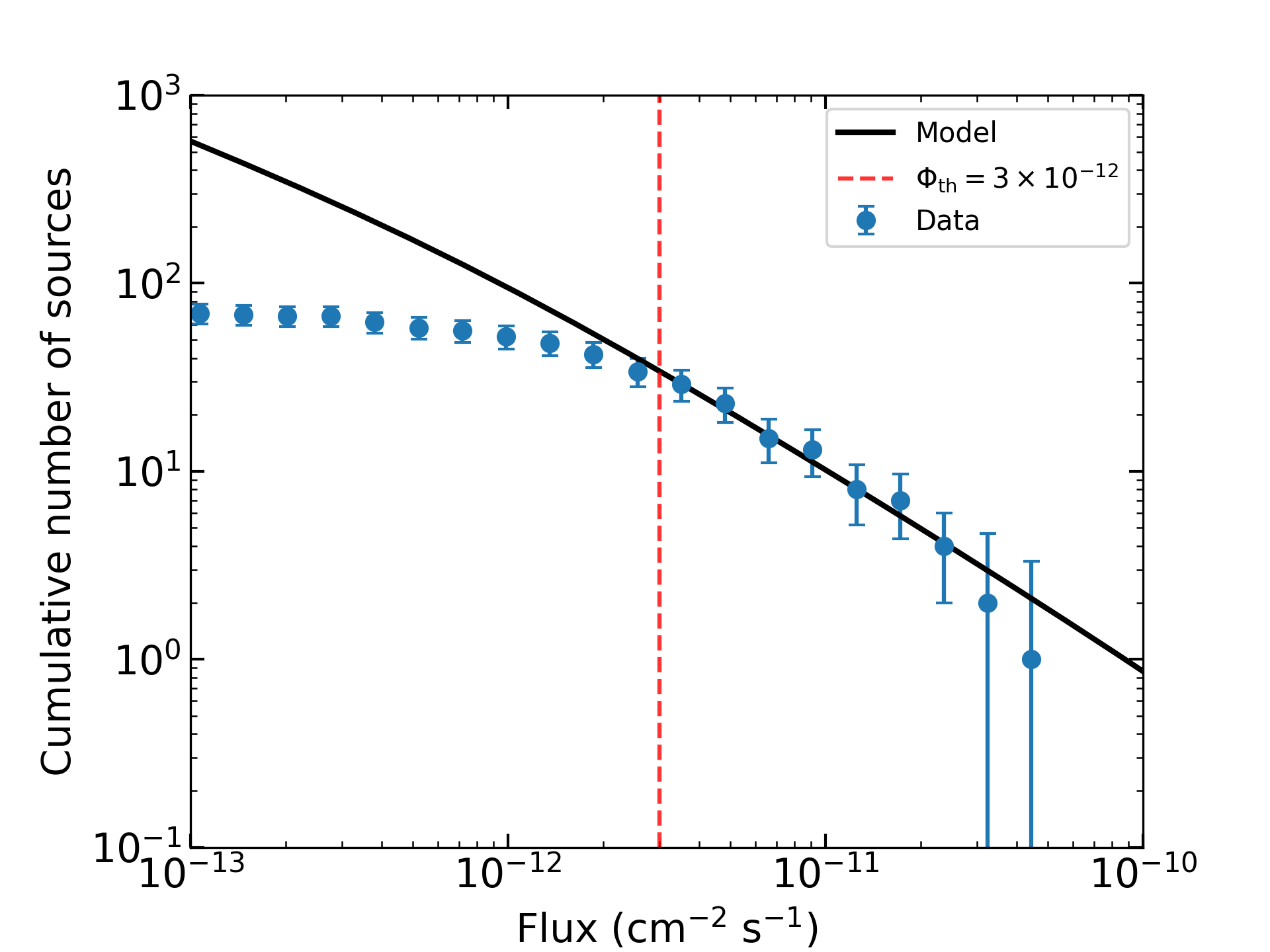}
\caption{Cumulative source number distribution $N(>\Phi)$ as a function of flux $\Phi$ for Galactic sources detected by LHAASO. The blue points with error bars represent the observed data, while the black line shows the best-fit model with parameters $\alpha=1.6$ and $L_{\rm cut}=10^{35}$~erg~s$^{-1}$. The red dashed line indicates the completeness flux threshold $\Phi_{\rm th}=3\times 10^{-12}$~cm$^{-2}$~s$^{-1}$, above which 29 sources are detected. Below this threshold, the observed source counts fall below the model prediction due to detection incompleteness.}
\label{fig:lognlogs}
\end{figure}

\begin{figure}[ht]
\centering
\includegraphics[width=0.6\textwidth]{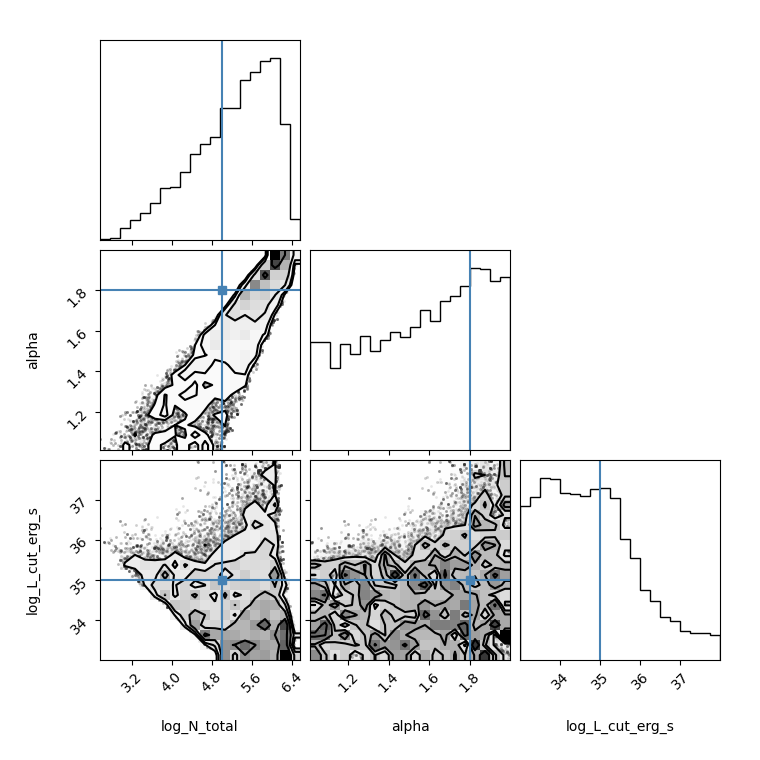}
\caption{Posterior probability distributions of the model parameters ($\log_{10} N$, $\alpha$, and $\log_{10}(L_{\rm cut}/{\rm erg\,s^{-1}})$) obtained from MCMC fitting to the observed cumulative source counts. The blue markers are the initial guess of the parameters.}
\label{fig:mcmc}
\end{figure}

\subsection{Fit to the LHAASO source distribution}

The model contains three free parameters: the total source number $N$, the power-law index $\alpha$, and the cutoff luminosity $L_{\rm cut}$. We constrain these parameters by fitting the predicted cumulative $N(>S)$ distribution to the LHAASO observed source counts using Markov Chain Monte Carlo (MCMC) sampling with \texttt{emcee} \cite{emcee2013}. The source count distribution for LHAASO is shown in Fig.~\ref{fig:lognlogs} as blue points. As can be seen from the figure, the detected sources can be considered complete at fluxes above approximately $\Phi = 3\times 10^{-12}~{\rm cm^{-2}~s^{-1}}$, so we set $\Phi_{\rm th}=3\times 10^{-12}~{\rm cm^{-2}~s^{-1}}$. There are 29 sources in total above $\Phi_{\rm th}$. We adopt flat priors: $\log_{10} N \in [0, 10]$, $\alpha \in [1.0, 2.0]$, and $\log_{10}(L_{\rm cut}/{\rm erg\,s^{-1}}) \in [33, 38]$. Fig.~\ref{fig:mcmc} shows the posterior distributions of the parameters obtained from the MCMC analysis.

As can be seen, due to the relatively small number of LHAASO sources, we cannot determine $\alpha$ very well. The MCMC posterior gives $\alpha=1.6^{+0.3}_{-0.4}$ and $\log_{10}(L_{\rm cut}/{\rm erg~s^{-1}}) = 34.6^{+1.2}_{-1.1}$. If we adopt $\alpha=1.6$ and $L_{\rm cut} = 10^{35}$~erg~s$^{-1}$ (corresponding to the black line in Fig.~\ref{fig:lognlogs}), we find that the total flux from unresolved sources amounts to $\sim41\%$ of the flux from resolved LHAASO sources. 
Therefore, the total flux from both resolved and unresolved point sources is $F_{\rm total} = F_{\rm res} + F_{\rm unres} \approx 1.4 \, F_{\rm res}$.
This result is consistent with previous analyses \cite{Cataldo2020ApJ,Lipari2025}. For instance, Lipari et al.~\cite{Lipari2025} performed a more detailed calculation of the flux ratio between unresolved and resolved sources for LHAASO data, obtaining a ratio of $f=F_{\text{unresolved}}/F_{\text{all}}=15-25\%$, which is generally in agreement with our result here.
This result indicates that even when accounting for unresolved sources, the total discrete source contribution remains insufficient to explain the observed diffuse neutrino emission from the Galactic disk.

As can be seen from Fig.~\ref{fig:lognlogs}, a somewhat smaller value of $\Phi_{\rm th}$ would also be compatible with the observations, but a smaller $\Phi_{\rm th}$ would lead to a shallower value of $\alpha$. For the conclusions of our paper, the current $\Phi_{\rm th}$ is conservative.

\section{Source Catalogs}\label{app catalogs}

Tables~\ref{tab:app_wcda} and \ref{tab:app_km2a} present the list of LHAASO sources used in this analysis. The columns include source name, right ascension (R.A.), declination (Decl.), angular extension (Ext.), spectral index ($\Gamma$), test statistic (TS), neutrino flux upper limit at the reference energy ($\Phi_{\nu}^{\rm UL}$), associated source name (Assoc.), source type, and whether the source is also detected by the other array (KM2A for WCDA sources, WCDA for KM2A sources). The TS and $\Phi_{\nu}^{\rm UL}$ columns are results derived in our analysis, while other columns are directly extracted from the 1LHAASO catalog.

\setlength{\tabcolsep}{3.5pt}
\startlongtable
\begin{deluxetable*}{cccccccccc}
\tablecaption{Single-source analysis results for 1LHAASO WCDA sources\label{tab:app_wcda}}
\tablehead{
\colhead{Source Name} &
\colhead{R.A.} &
\colhead{Decl.} &
\colhead{Ext.} &
\colhead{$\Gamma$} &
\colhead{TS} &
\colhead{$\Phi_{\nu,\rm 6TeV}^{\rm UL}$} &
\colhead{Assoc.} &
\colhead{Type} &
\colhead{KM2A}\\
\colhead{} & 
\colhead{[deg]} & 
\colhead{[deg]} & 
\colhead{[deg]} & 
\colhead{} & 
\colhead{} & 
\colhead{[$\rm cm^{-2}\,s^{-1}\,GeV^{-1}$]} & 
\colhead{} & 
\colhead{} & 
\colhead{}
}
\startdata
J0007+7303u  & 1.48   & 73.15  & $-$  & 2.74  & 3.84 & \num{8.57E-17} & CTA 1              & PWN                   &  \checkmark \\
J0056+6346u  & 13.78  & 63.96  & 0.33 & 2.35  & 0.91 & \num{4.15E-17} &                    &                       &  \checkmark \\
J0249+6022   & 41.52  & 60.49  & 0.71 & 2.52  & 0    & \num{5.98E-17} &                    &                       &  \checkmark \\
J0343+5254u* & 55.34  & 53.05  & 0.33 & 1.70  & 2.28 & \num{1.09E-17} & LHAASO J0341+5258  & UNID                  &  \checkmark \\
J0359+5406   & 59.68  & 54.21  & 0.22 & 1.74  & 0.25 & \num{1.40E-17} &                    &                       &  \checkmark \\
J0428+5531*  & 67.23  & 55.53  & 1.18 & 2.66  & 4.85 & \num{2.64E-16} &                    &                       &  \checkmark \\
J0500+4454   & 75.01  & 44.92  & 0.41 & 2.53  & 0.44 & \num{7.17E-17} &                    &                       &             \\
J0534+3533   & 83.38  & 35.48  & $-$  & 2.37  & 0    & \num{2.13E-17} &                    &                       &  \checkmark \\
J0534+2200u  & 83.62  & 22.01  & $-$  & 2.69  & 0.11 & \num{5.31E-17} & Crab               & PWN                   &  \checkmark \\
J0542+2311u  & 86.07  & 23.19  & 1.45 & 1.95  & 0    & \num{1.60E-17} & HAWC J0543+233     & TeV Halo              &  \checkmark \\
J0617+2234   & 94.35  & 22.57  & 0.59 & 2.92  & 0    & \num{3.28E-17} & IC 443             & Shell                 &             \\
J0622+3754   & 95.67  & 37.93  & 0.50 & 1.82  & 0    & \num{1.00E-17} & LHAASO J0621+3755  & PWN/TeV Halo          &  \checkmark \\
J0634+1741u  & 98.51  & 17.72  & 1.16 & 1.65  & 0    & \num{4.81E-18} & Geminga            & PWN/TeV Halo          &  \checkmark \\
J0703+1405   & 105.32 & 14.55  & 1.30 & 1.98  & 0    & \num{2.29E-17} & 2HWC J0700+143     & TeV Halo              &  \checkmark \\
J1809-1918u  & 272.66 & -19.32 & 0.35 & 2.24  & 0    & \num{1.64E-16} & HESS J1809-193     & UNID                  &  \checkmark \\
J1813-1245   & 273.35 & -12.73 & $-$  & 2.61  & 7.53 & \num{4.80E-16} & HESS J1813-126     & UNID                  &  \checkmark \\
J1814-1719u* & 273.69 & -17.33 & 0.71 & 2.83  & 0    & \num{4.01E-15} & 2HWC J1814-173     & UNID                  &  \checkmark \\
J1825-1418   & 276.29 & -14.32 & 0.81 & 2.98  & 0    & \num{5.47E-15} & HESS J1825-137     & PWN/TeV Halo          &  \checkmark \\
J1825-1256u  & 276.55 & -13.04 & 0.24 & 2.61  & 0    & \num{7.27E-16} & HESS J1826-130     & UNID                  &  \checkmark \\
J1825-1337u  & 276.55 & -13.73 & 0.17 & 2.55  & 0.29 & \num{4.65E-16} & HESS J1825-137     & PWN/TeV Halo          &  \checkmark \\
J1831-1007u* & 277.75 & -10.12 & 0.78 & 2.71  & 0.53 & \num{1.02E-15} & HESS J1831-098     & PWN                   &  \checkmark \\
J1834-0831   & 278.62 & -8.53  & 0.40 & 3.08  & 0.52 & \num{8.80E-16} & HESS J1834-087     & UNID                  &  \checkmark \\
J1837-0654u  & 279.39 & -6.90  & 0.34 & 2.92  & 0    & \num{2.24E-16} & HESS J1837-069     & PWN                   &  \checkmark \\
J1839-0548u  & 279.85 & -5.90  & 0.22 & 2.65  & 0.5  & \num{2.00E-16} & LHAASO J1839-0545  & UNID                  &  \checkmark \\
J1841-0519   & 280.33 & -5.33  & 0.60 & 2.88  & 0    & \num{9.74E-17} & HESS J1841-055     & UNID                  &  \checkmark \\
J1843-0335u  & 281.01 & -3.50  & 0.40 & 2.58  & 0    & \num{2.98E-17} & HESS J1843-033     & UNID                  &  \checkmark \\
J1848-0153u  & 282.06 & -1.89  & 0.51 & 2.65  & 0    & \num{9.34E-17} & HESS J1848-018     & Massive Star Cluster  &  \checkmark \\
J1850-0004u* & 282.74 & -0.07  & 0.46 & 2.49  & 0    & \num{1.01E-16} & HESS J1852-000     & UNID                  &  \checkmark \\
J1852+0050u* & 283.73 & 1.40   & 0.64 & 2.74  & 0    & \num{1.72E-16} & 2HWC J1852+013*    & UNID                  &  \checkmark \\
J1857+0245   & 284.37 & 2.75   & 0.24 & 2.93  & 0    & \num{8.55E-17} & HESS J1857+026     & UNID                  &             \\
J1857+0203u  & 284.50 & 1.98   & 0.19 & 2.46  & 0    & \num{4.24E-17} & HESS J1858+020     & UNID                  &  \checkmark \\
J1858+0330   & 284.79 & 3.70   & 0.52 & 2.63  & 1.1  & \num{1.38E-16} &                    &                       &  \checkmark \\
J1902+0648   & 285.58 & 6.80   & $-$  & 2.39  & 0    & \num{3.36E-17} &                    &                       &             \\
J1906+0712   & 286.56 & 7.20   & 0.21 & 2.72  & 0    & \num{1.21E-16} &                    &                       &             \\
J1907+0826   & 286.96 & 8.44   & 0.43 & 2.62  & 0    & \num{6.60E-17} & 2HWC J1907+084*    & UNID                  &             \\
J1908+0615u  & 287.05 & 6.26   & 0.43 & 2.42  & 3.99 & \num{1.20E-16} & MGRO J1908+06      & UNID                  &  \checkmark \\
J1910+0516*  & 287.88 & 5.07   & 0.29 & 2.54  & 0    & \num{8.84E-17} & SS433 w1           &                       &  \checkmark \\
J1912+1014u  & 288.22 & 10.25  & 0.36 & 2.68  & 2.53 & \num{1.25E-16} & HESS J1912+101     & Shell                 &  \checkmark \\
J1914+1150u  & 288.81 & 11.74  & 0.33 & 2.34  & 0.1  & \num{3.11E-17} & 2HWC J1914+117*    & UNID                  &  \checkmark \\
J1922+1403   & 290.70 & 14.06  & 0.18 & 2.62  & 0    & \num{2.28E-17} & W51                & SNR/Molec. Cloud      &  \checkmark \\
J1924+1609   & 291.09 & 16.15  & 1.45 & 2.54  & 0    & \num{1.35E-16} & 3HWC J1923+169     & UNID                  &  \checkmark \\
J1928+1746u  & 292.14 & 17.78  & 0.17 & 2.22  & 1.17 & \num{7.16E-17} & 2HWC J1928+177     & UNID                  &  \checkmark \\
J1929+1846u* & 292.34 & 18.77  & 0.49 & 2.37  & 0.63 & \num{7.98E-17} & SNR G054.1+00.3    & PWN                   &  \checkmark \\
J1937+2128   & 294.30 & 21.00  & 1.25 & 2.43  & 0.49 & \num{1.36E-16} & 3HWC J1935+213     & UNID                  &  \checkmark \\
J1945+2424*  & 296.36 & 24.40  & 1.29 & 2.56  & 4.02 & \num{2.14E-16} & 2HWC J1949+244     & UNID                  &  \checkmark \\
J1952+2922   & 298.05 & 29.38  & $-$  & 2.52  & 0    & \num{3.08E-17} & 2HWC J1953+294     & PWN                   &             \\
J1954+2836u  & 298.50 & 28.57  & $-$  & 2.22  & 1.39 & \num{3.40E-17} & 2HWC J1955+285     & UNID                  &  \checkmark \\
J1954+3253   & 298.63 & 32.88  & 0.17 & 2.45  & 1.51 & \num{3.40E-17} &                    &                       &             \\
J1956+2921   & 299.24 & 29.35  & 0.99 & 2.03  & 5.69 & \num{4.87E-17} & LHAASO J1956+2845  & UNID                  &  \checkmark \\
J2002+3244u  & 300.64 & 32.74  & $-$  & 2.21  & 4.42 & \num{2.17E-17} &                    &                       &  \checkmark \\
J2005+3415*  & 301.30 & 34.25  & 0.74 & 2.58  & 0    & \num{1.18E-16} & 2HWC J2006+341     & UNID                  &  \checkmark \\
J2005+3050   & 301.37 & 30.99  & 0.21 & 1.99  & 0    & \num{1.97E-17} & 3HWC J2005+311     & UNID                  &  \checkmark \\
J2018+3643u  & 304.61 & 36.75  & 0.26 & 1.94  & 0.76 & \num{1.72E-17} & MGRO J2019+37      & PWN                   &  \checkmark \\
J2020+4034   & 305.03 & 40.57  & 0.38 & 2.91  & 0    & \num{2.16E-17} & VER J2019+407      & UNID                  &  \checkmark \\
J2020+3638   & 305.14 & 36.63  & 1.27 & 2.62  & 3.87 & \num{1.05E-16} & VER J2019+368      & UNID                  &             \\
J2020+3649u  & 305.25 & 36.91  & 0.15 & 1.78  & 0    & \num{7.76E-18} & VER J2019+368      & UNID                  &  \checkmark \\
J2031+4052u* & 307.90 & 40.88  & 0.25 & 2.81  & 8.48 & \num{1.78E-16} & LHAASO J2032+4102  & UNID                  &  \checkmark \\
J2031+4127u  & 307.85 & 41.60  & 0.27 & 2.29  & 2.76 & \num{7.47E-17} & PSR J2032+4127     & Binary                &  \checkmark \\
J2108+5153u  & 316.83 & 51.95  & 0.14 & 1.44  & 2.43 & \num{4.19E-18} & LHAASO J2108+5157  & DARK                  &  \checkmark \\
J2200+5643u  & 330.38 & 56.73  & 0.43 & 1.77  & 1.08 & \num{1.20E-17} &                    &                       &  \checkmark \\
J2228+6100u  & 336.79 & 61.02  & 0.25 & 2.26  & 0    & \num{2.46E-17} & SNR G106.3+02.7    & Shell                 &  \checkmark \\
J2229+5927u  & 337.26 & 59.45  & 1.98 & 2.67  & 3.56 & \num{2.15E-16} &                    &                       &  \checkmark \\
J2238+5900   & 339.40 & 58.92  & 0.51 & 2.39  & 0    & \num{5.56E-17} &                    &                       &  \checkmark \\
J2323+5854   & 350.80 & 58.90  & $-$  & 3.18  & 1.53 & \num{5.02E-17} & Cassiopeia A       & Shell                 &             \\
\enddata

\end{deluxetable*}

\startlongtable
\begin{deluxetable*}{cccccccccc}
\tablecaption{Single-source analysis results for 1LHAASO KM2A sources\label{tab:app_km2a}}
\tablehead{
\colhead{Source Name} &
\colhead{R.A.} &
\colhead{Decl.} &
\colhead{Ext.} &
\colhead{$\Gamma$} &
\colhead{TS} &
\colhead{$\Phi_{\nu,\rm 25TeV}^{\rm UL}$} &
\colhead{Assoc.} &
\colhead{Type} &
\colhead{WCDA}\\
\colhead{} & 
\colhead{[deg]} & 
\colhead{[deg]} & 
\colhead{[deg]} & 
\colhead{} & 
\colhead{} & 
\colhead{[$\rm cm^{-2}\,s^{-1}\,GeV^{-1}$]} & 
\colhead{} & 
\colhead{} & 
\colhead{}
}
\startdata
J0007+5659u  & 1.86   & 57.00  & $-$  & 3.10 & 0.89 & \num{3.24E-19} &                   &                      &            \\
J0007+7303u  & 1.91   & 73.07  & 0.17 & 3.40 & 3.46 & \num{3.30E-19} & CTA 1             & PWN                  & \checkmark \\
J0056+6346u  & 14.10  & 63.77  & 0.24 & 3.33 & 2.02 & \num{3.48E-19} &                   &                      & \checkmark \\
J0206+4302u  & 31.70  & 43.05  & $-$  & 2.62 & 0    & \num{7.65E-19} &                   &                      &            \\
J0212+4254u  & 33.01  & 42.91  & $-$  & 2.45 & 0.54 & \num{8.30E-19} &                   &                      &            \\
J0216+4237u  & 34.10  & 42.63  & $-$  & 2.58 & 0    & \num{4.96E-19} &                   &                      &            \\
J0249+6022   & 42.39  & 60.37  & 0.38 & 3.82 & 0.66 & \num{4.16E-20} &                   &                      & \checkmark \\
J0339+5307   & 54.79  & 53.13  & $-$  & 3.64 & 0.26 & \num{6.89E-20} & LHAASO J0341+5258 & UNID                 &            \\
J0343+5254u* & 55.79  & 52.91  & 0.20 & 3.53 & 3.05 & \num{2.17E-19} & LHAASO J0341+5258 & UNID                 & \checkmark \\
J0359+5406   & 59.78  & 54.10  & 0.30 & 3.84 & 0.55 & \num{3.96E-20} &                   &                      & \checkmark \\
J0428+5531*  & 66.63  & 54.63  & 0.32 & 3.45 & 4    & \num{3.50E-19} &                   &                      & \checkmark \\
J0534+3533   & 83.53  & 35.56  & $-$  & 4.89 & 0    & \num{1.50E-22} &                   &                      & \checkmark \\
J0534+2200u  & 83.61  & 22.04  & $-$  & 3.19 & 0.14 & \num{3.45E-19} & Crab              & PWN                  & \checkmark \\
J0542+2311u  & 85.71  & 23.20  & 0.98 & 3.74 & 0    & \num{6.61E-20} & HAWC J0543+233    & TeV Halo             & \checkmark \\
J0622+3754   & 95.50  & 37.90  & 0.46 & 3.68 & 0    & \num{4.60E-20} & LHAASO J0621+3755 & PWN/TeV Halo         & \checkmark \\
J0631+1040   & 97.77  & 10.67  & $-$  & 3.33 & 0.18 & \num{2.60E-19} & 3HWC J0631+107    & UNID                 &            \\
J0634+1741u  & 98.57  & 17.69  & 0.89 & 3.69 & 0.92 & \num{1.71E-19} & Geminga           & PWN/TeV Halo         & \checkmark \\
J0635+0619   & 98.76  & 6.33   & 0.60 & 3.67 & 9.37 & \num{4.51E-19} & HAWC J0635+070    & TeV Halo             &            \\
J0703+1405   & 105.83 & 14.10  & 1.88 & 3.98 & 0    & \num{1.00E-19} & 2HWC J0700+143    & TeV Halo             & \checkmark \\
J1809-1918u  & 272.38 & -19.30 & $-$  & 3.51 & 0    & \num{1.85E-16} & HESS J1809-193    & UNID                 & \checkmark \\
J1813-1245   & 273.36 & -12.75 & $-$  & 3.66 & 7.54 & \num{1.85E-16} & HESS J1813-126    & UNID                 & \checkmark \\
J1814-1719u* & 273.27 & -17.89 & $-$  & 3.49 & 0.16 & \num{3.80E-16} & 2HWC J1814-173    & UNID                 & \checkmark \\
J1814-1636u  & 273.72 & -16.62 & 0.68 & 3.74 & 0    & \num{2.31E-16} & 2HWC J1814-173    & UNID                 &            \\
J1825-1418   & 276.25 & -14.00 & 0.81 & 3.53 & 0    & \num{1.60E-16} & HESS J1825-137    & PWN/TeV Halo         & \checkmark \\
J1825-1256u  & 276.44 & -12.94 & $-$  & 3.33 & 0    & \num{4.42E-17} & HESS J1826-130    & UNID                 & \checkmark \\
J1825-1337u  & 276.45 & -13.63 & $-$  & 3.28 & 0.1  & \num{6.32E-17} & HESS J1825-137    & PWN/TeV Halo         & \checkmark \\
J1831-1007u* & 277.81 & -9.83  & 0.26 & 3.30 & 0    & \num{6.34E-18} & HESS J1831-098    & PWN                  & \checkmark \\
J1831-1028   & 277.84 & -10.48 & 0.94 & 3.53 & 0    & \num{6.51E-17} & HESS J1833-105    & PWN                  &            \\
J1834-0831   & 278.44 & -8.38  & 0.40 & 3.63 & 0.34 & \num{1.63E-18} & HESS J1834-087    & UNID                 & \checkmark \\
J1837-0654u  & 279.31 & -6.86  & 0.33 & 3.70 & 0    & \num{2.85E-19} & HESS J1837-069    & PWN                  & \checkmark \\
J1839-0548u  & 279.79 & -5.81  & 0.22 & 3.24 & 0.09 & \num{1.17E-18} & LHAASO J1839-0545 & UNID                 & \checkmark \\
J1841-0519   & 280.21 & -5.23  & 0.62 & 3.85 & 0    & \num{9.08E-20} & HESS J1841-055    & UNID                 & \checkmark \\
J1843-0335u  & 280.91 & -3.60  & 0.36 & 3.44 & 0    & \num{2.18E-19} & HESS J1843-033    & UNID                 & \checkmark \\
J1848-0153u  & 282.02 & -1.78  & 0.56 & 3.69 & 0.14 & \num{1.32E-19} & HESS J1848-018    & Massive Star Cluster & \checkmark \\
J1848-0001u  & 282.19 & -0.02  & $-$  & 2.75 & 0    & \num{1.17E-18} & IGR J18490-0000   & PWN                  &            \\
J1850-0004u* & 282.89 & -0.07  & 0.21 & 3.15 & 0    & \num{6.17E-19} & HESS J1852-000    & UNID                 & \checkmark \\
J1852+0050u* & 283.10 & 0.84   & 0.85 & 3.64 & 0    & \num{4.28E-19} & 2HWC J1852+013*   & UNID                 & \checkmark \\
J1857+0203u  & 284.38 & 2.06   & 0.28 & 3.31 & 0    & \num{4.96E-19} & HESS J1858+020    & UNID                 & \checkmark \\
J1858+0330   & 284.59 & 3.51   & 0.43 & 3.78 & 0.31 & \num{8.20E-20} &                   &                      & \checkmark \\
J1908+0615u  & 287.05 & 6.26   & 0.36 & 2.82 & 4.3  & \num{2.11E-18} & MGRO J1908+06     & UNID                 & \checkmark \\
J1910+0516*  & 287.55 & 5.28   & $-$  & 3.15 & 2.2  & \num{8.31E-19} & SS 433 w1         &                      & \checkmark \\
J1912+1014u  & 288.38 & 10.50  & 0.5  & 3.26 & 1.52 & \num{7.59E-19} & HESS J1912+101    & Shell                & \checkmark \\
J1913+0501   & 288.28 & 5.03   & $-$  & 3.3  & 1.61 & \num{5.01E-19} & SS 433 e1         &                      &            \\
J1914+1150u  & 288.73 & 11.84  & 0.21 & 3.41 & 0    & \num{1.95E-19} & 2HWC J1914+117*   & UNID                 & \checkmark \\
J1919+1556   & 289.78 & 15.93  & $-$  & 4.71 & 0    & \num{1.16E-21} & 3HWC J1918+159    & UNID                 &            \\
J1922+1403   & 290.73 & 14.11  & $-$  & 3.79 & 0    & \num{1.78E-20} & W 51              & SNR/Molec. Cloud     & \checkmark \\
J1924+1609   & 290.53 & 15.71  & 1.22 & 3.61 & 0    & \num{2.00E-19} & 3HWC J1923+169    & UNID                 & \checkmark \\
J1928+1813u  & 292.07 & 18.23  & 0.63 & 3.24 & 1.5  & \num{6.20E-19} & 2HWC J1928+177    & UNID                 &            \\
J1928+1746u  & 292.17 & 17.89  & $-$  & 3.1  & 1.28 & \num{5.91E-19} & 2HWC J1928+177    & UNID                 & \checkmark \\
J1929+1846u* & 292.04 & 18.97  & $-$  & 3.11 & 0.75 & \num{5.81E-19} & SNR G054.1+00.3   & PWN                  & \checkmark \\
J1931+1653   & 292.79 & 16.90  & $-$  & 3.15 & 4.89 & \num{1.08E-18} &                   &                      &            \\
J1937+2128   & 294.32 & 21.48  & 1.43 & 3.4  & 0.84 & \num{6.20E-19} & 3HWC J1935+213    & UNID                 & \checkmark \\
J1945+2424*  & 297.42 & 23.97  & 0.36 & 3.93 & 0.5  & \num{3.63E-20} & 2HWC J1949+244    & UNID                 & \checkmark \\
J1951+2608   & 297.94 & 26.15  & 1.0  & 3.43 & 3.68 & \num{7.20E-19} & 3HWC J1951+266    & UNID                 &            \\
J1954+2836u  & 298.55 & 28.60  & $-$  & 2.92 & 1.59 & \num{1.20E-18} & 2HWC J1955+285    & UNID                 & \checkmark \\
J1956+2921   & 298.84 & 28.92  & 0.78 & 3.42 & 3.96 & \num{6.36E-19} & LHAASO J1956+2845 & UNID                 & \checkmark \\
J1959+2846u  & 299.78 & 28.78  & 0.29 & 2.9  & 0.5  & \num{1.08E-18} &                   &                      &            \\
J1959+1129u  & 299.82 & 11.49  & $-$  & 2.69 & 0.61 & \num{7.51E-19} &                   &                      &            \\
J2002+3244u  & 300.60 & 32.64  & $-$  & 2.7  & 6.24 & \num{1.70E-18} &                   &                      & \checkmark \\
J2005+3415*  & 301.81 & 33.87  & 0.33 & 3.79 & 0.08 & \num{4.07E-20} & 2HWC J2006+341    & UNID                 & \checkmark \\
J2005+3050   & 301.45 & 30.85  & 0.27 & 3.62 & 0    & \num{6.82E-20} & 3HWC J2005+311    & UNID                 & \checkmark \\
J2018+3643u  & 304.65 & 36.72  & 0.24 & 3.46 & 0.68 & \num{1.88E-19} & MGRO J2019+37     & PWN                  & \checkmark \\
J2020+4034   & 305.20 & 40.43  & $-$  & 3.56 & 0.15 & \num{7.33E-20} & VER J2019+407     & UNID                 & \checkmark \\
J2020+3649u  & 305.23 & 36.82  & 0.12 & 3.31 & 0    & \num{1.75E-19} & VER J2019+368     & UNID                 & \checkmark \\
J2027+3657   & 306.88 & 36.95  & 0.38 & 3.21 & 0.06 & \num{4.79E-19} &                   &                      &            \\
J2028+3352   & 307.21 & 33.88  & 1.7  & 3.38 & 0    & \num{7.29E-19} &                   &                      &            \\
J2031+4052u* & 308.14 & 40.88  & $-$  & 2.13 & 5.33 & \num{2.88E-18} & LHAASO J2032+4102 & UNID                 & \checkmark \\
J2031+4127u  & 307.95 & 41.46  & 0.22 & 3.45 & 4.48 & \num{3.58E-19} & PSR J2032+4127    & Binary               & \checkmark \\
J2047+4434   & 311.92 & 44.58  & 0.42 & 3.17 & 0    & \num{2.72E-19} &                   &                      &            \\
J2108+5153u  & 317.10 & 51.90  & 0.19 & 2.97 & 2.13 & \num{1.29E-18} & LHAASO J2108+5157 & DARK                 & \checkmark \\
J2200+5643u  & 330.08 & 56.73  & 0.54 & 3.44 & 0    & \num{2.70E-19} &                   &                      & \checkmark \\
J2228+6100u  & 337.01 & 61.00  & 0.35 & 2.95 & 0    & \num{4.38E-19} & SNR G106.3+02.7   & Shell                & \checkmark \\
J2229+5927u  & 337.88 & 59.55  & 1.74 & 3.53 & 0    & \num{5.21E-19} &                   &                      & \checkmark \\
J2238+5900   & 339.54 & 59.00  & 0.43 & 3.55 & 0    & \num{1.22E-19} &                   &                      & \checkmark \\
\enddata

\end{deluxetable*}

\end{appendix}
\end{document}